\documentclass[journal]{IEEEtran}
\usepackage{jdhmacros}
\usepackage{algorithm}
\usepackage{algorithmic, bm, textcomp}

\newcommand{\RN}[1]{%
  \textup{\uppercase\expandafter{\romannumeral#1}}%
}

\newcommand{\jdh}[1]{}
\renewcommand{\jdh}[1]{{\color{magenta}{#1}}} 

\usepackage[colorlinks]{hyperref}

\begin{document}

\title{\huge Improved Support Recovery Guarantees for the Group Lasso\\
With Applications to Structural Health Monitoring}

\author{Mojtaba Kadkhodaie Elyaderani,~\IEEEmembership{{Student Member},~IEEE}, Swayambhoo Jain,~\IEEEmembership{{Student Member},~IEEE},\\Jeffrey Druce,
Stefano Gonella, and Jarvis Haupt,~\IEEEmembership{{Senior} Member,~IEEE}\thanks{Submitted August 28, 2017.  Restructured October 2, 2017 and October 16, 2017. Revised May 14, 2018.}\thanks{MKE and JH are with the Department of Electrical and Computer Engineering at the University of Minnesota -- Twin Cities. Tel/fax: (612) 625-3300 \ /\ (612) 625-4583. Emails: {\tt \{kadkh004, jdhaupt\}@umn.edu}.  {SJ is with Technicolor Los Altos Research Center, Los Altos, CA. Email: {\tt jainx174@umn.edu}}. {JD is with Charles River Analytics, Cambridge MA. Email:{\tt jdruce@cra.com}.}. SG is with the Department of Civil, Environmental, and Geo- Engineering at the University of Minnesota -- Twin Cities. Tel/fax: (612) 625-5522 \ /\ (612) 626-7750. Email: {\tt sgonella@umn.edu}.}\thanks{Preliminary versions of this work were submitted to CAMSAP 2015 and ICASSP 2017. The authors graciously acknowledge support from the University of Minnesota Digital Technology Center, NSF Award No. CCF-1217751, and the DARPA Young Faculty Award, Grant N66001-14-1-4047.}}

\maketitle

\begin{abstract}
This paper considers the problem of estimating an unknown high dimensional signal from noisy linear measurements, {when} the signal is assumed to possess a \emph{group-sparse} structure in a {known,} fixed dictionary.  We consider signals generated according to a natural probabilistic model, and establish new conditions under which the set of indices of the non-zero groups of the signal (called the group-level support) may be accurately estimated via the group Lasso.  Our results strengthen existing coherence-based analyses that exhibit the well-known ``square root'' bottleneck, allowing for the number of recoverable nonzero groups to be nearly as large as the total number of groups.  We also establish a sufficient recovery condition relating the number of nonzero groups and the signal to noise ratio (quantified in terms of the ratio of the squared Euclidean norms of nonzero groups and the variance of the random additive {measurement} noise), and evaluate this trend empirically.  Finally, we examine the implications of our results in the context of a structural health monitoring application, where the group Lasso approach facilitates demixing of a propagating acoustic wavefield, acquired on the material surface by a scanning laser Doppler vibrometer, into antithetical components, one of which indicates the locations of internal material defects. 
\end{abstract}
\vspace{-0.0in}
\begin{IEEEkeywords}
anomaly detection, convex demixing, group Lasso, non-destructive evaluation, primal-dual witness, support recovery
\end{IEEEkeywords} 


\section{Introduction}
In recent years, the recovery of structured signals from noisy linear measurements has been an active area of research in the fields of signal processing, high-dimensional statistics, and machine learning \cite{donoho2006compressed,candes2009exact,negahban2009unified,chandrasekaran2012convex}. 
Suppose an unknown signal $\bbeta^*\in\RR^p$ is observed {via} the noisy linear measurement model 
\begin{equation}\label{eq:obs-model}
\by = \bX \bbeta^* +\bw,
\end{equation}
where $\by\in \RR^n$ is the vector of observations, $\bX\in \RR^{n\times p}$ is the dictionary matrix, and $\bw\in\RR^n$ {describes} noise and/or model inaccuracies.  Many contemporary works assume $n<p$, in which case it is (in general) impossible to uniquely recover general $\bbeta^*$ from the measurements. However, exploiting the fact that in {many} applications the signal of interest exhibits a low-dimensional structure opens the opportunity for using contemporary inference approaches from high dimensional statistics and compressed sensing. 
Remarkable results such as those established in \cite{tropp2008conditioning, candes2009near} illustrate that, when the signal of interest is sparse and the dictionary $\bX$ satisfies certain {structural} conditions, one can accurately infer $\bbeta^*$ by solving the so-called Lasso problem \cite{tibshirani1996regression} even when the number of non-zero entries of $\bbeta^*$ is {nearly} proportional to the number of measurements.

Here we consider settings where the signal of interest is group-sparse,  meaning that given a partition of its entries into groups only a few are non-zero. In these settings, the  group Lasso estimator, 
\begin{equation}\label{GLasso}
\hat{\bbeta}=\arg\min_{\bbeta \in \RR^{p}} \frac{1}{2}||\by-\bX\bbeta||_2^2+\sum_{g=1}^{G}\lambda_g ||\bbeta_{\cI_g}||_2,
\end{equation}
originally proposed in  \cite{yuan2006model} is a natural approach to infer the unknown signal.  In {the formulation of interest here}, $\bbeta$ is expressed in terms of a  given (known, fixed) partition of its entries into $G$ {non-overlapping} blocks or groups
\begin{equation}\label{eq:grouping}
\bbeta = \left[ (\bbeta_{\cI_1})^T (\bbeta_{\cI_2})^T \cdots (\bbeta_{\cI_G})^T\right]^T,
\end{equation}
 where $\bbeta_{\cI_g} \in \RR^{d_g}$ represents the $g$-th constituent block of $\bbeta$ with $\cI_g$ denoting the (possibly non-contiguous) subset of entries of $\bbeta$ that belong to the $g$-th block, {$d_g$ denotes the cardinality of the $g$-th block}, the $\lambda_g>0$ are regularization parameters, and $\|\cdot\|_2$ denotes the Euclidean norm. This estimator exploits the extra knowledge about the natural grouping of the signal entries by attempting to enforce that only a few possible groups be present in the estimate.  When group structure is present, its performance {can exceed} that of the standard Lasso estimator, where each element of $\bbeta$ is a singleton group (see, e.g., \cite{yuan2006model, bach2008consistency}). 
 
Our motivation here comes from a laser-enabled structural health monitoring application (described in detail in the sequel), where a material under test is subjected to a propagating acoustic wavefield, and its spatiotemporal displacement response is measured on its surface by a non-contact scanning laser Doppler vibrometer.  We posit that over several consecutive time steps, the propagating wavefield may be well approximated as a sum of two antithetical components -- one that models the propagating wavefield in the bulk (undamaged) portion of the medium, and another that captures the local (subtle) perturbations that arise in the wavefield in the neighborhood(s) of material defects -- each of which may be expressed in terms of a group-sparse signal in an appropriately chosen (fixed) spatiotemporal dictionary.  Under this model, accurate localization of the defected regions amounts to correctly identifying the locations of the nonzero groups that comprise the corresponding component of the spatiotemporal wavefield.  For this we employ the group Lasso estimator, and thus aim to explore and quantify its performance for identifying the locations of nonzero groups (called \emph{support recovery} in the literature) from generally noisy observations, in finite-sample regimes, in settings where the dictionary is deterministic and fixed.

\subsection{Prior Work and Our Contributions}
 
To date, there are numerous works that provide various forms of analytical guarantees for the group Lasso (e.g., \cite{yuan2006model, bach2008consistency, tropp2006algorithms2, nardi2008asymptotic, meier2008group, liu2009estimation, huang2010benefit, obozinski2011support, kolar2011union, fang2011sparse, lounici2011oracle, vaiter2012degrees, bajwa2015conditioning, ahsen2017}) and related estimation problems (e.g., \cite{cotter2005sparse, tropp2006algorithms1, gribonval2008atoms, stojnic2009reconstruction, eldar2009robust, eldar2010average, eldar2010block, baraniuk2010model, stojnic2010ell, boufounos2011sparse, fang2011recovery, ben2011near, kim2012compressive, davies2012rank, lee2012subspace, elhamifar2012block, rao2012universal, duarte2013measurement}) for block-sparse signal recovery, under a number of different modeling and structural assumptions.  Below we provide a concise, high-level summary of these existing works, highlighting specifically the modeling assumptions they employ and the nature of the guarantees they provide.  We then outline our contributions within this context. 

Utilizing standard tools from convex analysis, the analytical guarantees presented in most of the existing group Lasso analyses referenced above rely upon conditions of the dictionary matrices that are combinatorial to test (e.g., the so-called \emph{irreducibility} conditions, restricted eigenvalue conditions, or restricted isometry conditions).  To our knowledge, the lone exception is \cite{bajwa2015conditioning}, whose analyses rely upon easily verifiable coherence conditions of the dictionary matrix (and employ randomized signal models to avoid the ``square root bottleneck''  described later in more detail).  Indeed, one common, tractable analytical approach to alleviating the burden associated with validating combinatorial  conditions on the dictionary matrix is to assume it is randomly generated (e.g., as in \cite{bach2008consistency, meier2008group, obozinski2011support, kolar2011union}).  This is contrary to our setup, as we aim to use fixed dictionaries, and aim for easily verifiable analytical conditions on them; hence, our analysis builds upon the framework introduced in \cite{bajwa2015conditioning}.

In terms of the type of recovery guarantees provided, several of the aforementioned works do indeed consider support recovery performance of the group Lasso \cite{tropp2006algorithms2, bach2008consistency, nardi2008asymptotic, obozinski2011support, kolar2011union}.  Some of these provide asymptotic analyses (e.g., \cite{bach2008consistency, nardi2008asymptotic}), while our interest here is on finite sample guarantees. More generally, all of these existing works again rely on dictionary conditions that are combinatorial to verify for a given dictionary. Again, the sole exception we are aware of is \cite{bajwa2015conditioning}, though that work only examines regression problems, providing Euclidean prediction error estimates using their coherence-based analyses (and exact recovery guarantees in noise-free settings, which are not our interest here).

We briefly comment on the related group-sparse signal recovery works cited above. Several of the earliest of these works aim to solve the simultaneous sparse approximation problem (also called the multiple measurement vectors problem), and to that end, examine the performance of greedy algorithms such as matching pursuit and its variants (e.g., \cite{cotter2005sparse, tropp2006algorithms1, gribonval2008atoms}).  Matching pursuit variants were also explored in \cite{fang2011recovery, ben2011near, kim2012compressive, davies2012rank}.  Others are focused primarily on noise-free settings, and seek to identify sufficient \cite{stojnic2009reconstruction, eldar2010average, eldar2010block, boufounos2011sparse, elhamifar2012block, rao2012universal, duarte2013measurement} (and in some cases, necessary \cite{stojnic2010ell, boufounos2011sparse}) conditions for block-sparse signal recovery.  It is worth noting that several of these works (e.g., \cite{eldar2010average, eldar2010block, lv2011group}) do indeed employ notions of block coherence in their analyses which, in settings where both the dictionary matrix and the signal are assumed deterministic, exhibit the ``square root'' bottleneck outlined below.  We offer some more quantitative comparisons between our main results and this line of work in the sequel.  

In this work, we make the following contributions:
\begin{itemize}
\item We derive new support recovery guarantees for the group Lasso for settings characterized by noisy observations and fixed (deterministic) dictionaries, establishing sufficient conditions for exact support recovery in which the number of recoverable nonzero groups can be nearly as large as the ratio between the number of measurements ($n$, in the notation above) and the maximum group size, up to constant and logarithmic factors.  This improves substantially upon existing deterministic coherence-based analyses that exhibit the well-known \emph{square root bottleneck}, where the sufficient conditions for recovery prescribe the number of nonzero blocks be, \emph{in best-case scenarios}, no larger than (constant and logarithmic factors times) the square root of the aforementioned ratio. We accomplish this by employing a mild statistical prior on the signals of interest, and leveraging (in part) powerful recent results quantifying the coherence of random \emph{block} subdictionaries of fixed dictionaries \cite{bajwa2015conditioning}.  
\item For the same scenario, we identify (analytically) intrinsic relations that quantify support recoverability as a function of the interplay between the number of nonzero groups and their magnitudes, quantified in terms of the groups' Euclidean norms.  This elucidates a tradeoff between rarity and weakness of defects that are detectable in our structural health monitoring application of interest.  We evaluate our analytical predictions through numerical simulation on both synthetic data adhering to our generative model and application data generated by finite element simulation.
\end{itemize}


In terms of the motivating application itself, our investigation builds upon and expands our own previous efforts along these lines, which include defect localization methods based on dictionary learning (for settings where the bases representing the antithetical components are not a priori fixed, but instead are learned from the data itself), \cite{druce2015anomaly}, modeling analysis and experimental investigations of the efficacy of the group Lasso method (using a priori fixed dictionaries) for defect localization \cite{druce2015structural, druce2016defect, druce2017locating}, a preliminary analysis of the group Lasso for this application \cite{kadkhodaie2015locating}, and a conference-length summary of the results of the present work \cite{elyaderani2017group}. We provide a brief background (with selected references) for our motivating application in Sec.~\ref{sec:prb-frml}.

\subsection{Notation and Organization}
Throughout the paper, bold-face lowercase and uppercase letters will be used to denote vectors and matrices, respectively.  For a vector $\bv$, we use $\|\bv\|_2$ to denote its Euclidean norm and  for a matrix $\bV$, its spectral and Frobenius norms are denoted by $\|\bV\|_{2\rightarrow 2}$ and $\|\bV\|_F$, respectively. Moreover, the sum of the absolute values of the entries of a matrix $\bV$ (or a vector $\bv$) are denoted by $\|\bV\|_1$ (or $\|\bv\|_1$) and the maximum absolute value of entries is represented by $\|\bV\|_{\infty}$ (or $\|\bv\|_{\infty}$).

For any integer $m>0$, we use $[m]$ as the shorthand for the set $\{1,2,\cdots,m\}$. If $n$ denotes the length of $\bbeta$ and the number of columns of $\bX$, then for the index set $\cI_g \subset [n]$, $\bbeta_{\cI_g}$ will denote the group of entries of $\bbeta$ whose indices belong to this set and $\bX_{\cI_g}$ will be the {submatrix comprised of} columns of $\bX$ indexed by $\cI_g$. For a column-wise block partitioned matrix $\bM=\left[\bM_{\cI_1}\, \bM_{\cI_2} \cdots \bM_{\cI_G} \right]$ the norm $\|\bM\|_{B,1}$  is defined as $$\|\bM\|_{B,1} := \max_{g\in[G]} \|\bM_{\cI_g}\|_{2\rightarrow 2}.$$

Throughout the paper, we will use different notions of signal support defined as follows:
\begin{itemize}
\item $\cS(\bbeta) := \{j\in [n]: \beta_j \neq 0\}$ will be the support of $\bbeta\in \RR^n$.
\item $\cG(\bbeta) := \{g\in[G]: \bbeta_{\cI_g} \neq \mathbf{0}\}$ will denote the set that contains the indices of the nonzero groups of $\bbeta$, where $G$ is the total number of groups. 
\item $\cS_{\cG}(\bbeta) := \cup_{g\in \cG(\bbeta)} \cI_{g}$. In words, $\cS_{\cG}(\bbeta)$ will denote the set that contains all indices comprising groups that are nonzero (even if there are zero elements at those particular indices).  Note that $\cS(\bbeta) \subseteq \cS_{\cG}(\bbeta)$.
\end{itemize} 

We let $d_{\min} := \min_{g \in [G]}d_g  \ \mbox{ and } \ d_{\max} := \max_{g\in [G]}d_g$ be the minimum and maximum group sizes, respectively, and $$d_{\cG(\bbeta)} := \sum_{g\in \cG(\bbeta)} d_g$$ be the total number of entries in the group-level support $\cG(\bbeta)$ of $\bbeta$. Similarly, we define $$\lambda_{\rm min} :=\min_{g\in [G]} \lambda_g \ \mbox{ and } \ \lambda_{\rm max} := \max_{g \in [G]} \lambda_g$$ to be the minimum and maximum regularization constants, respectively, and let $\blambda_{\cG(\bbeta)}$ be the $|\cG(\bbeta)|$-dimensional vector whose entries are the regularization parameters corresponding to the groups in $\cG(\bbeta^*)$.
In order to {clarify} notation, we will use $\cG^*$, $\cS^*_{\cG}$, and $d^*_{\cG}$ as  abbreviations for $\cG(\bbeta^*)$, $\cS_{\cG}(\bbeta^*)$, and $d_{\cG(\bbeta^*)}$, respectively.

{

The rest of the paper is organized as follows. We provide our main recovery result in Section~\ref{sec:main-result}. The implications of our analysis in the context of the structural health monitoring application is discussed in Section~\ref{sec:prb-frml}. We validate our theoretical results experimentally in Section~\ref{sec:exp-results}.
Section~\ref{sec:main-proof} outlines the main steps of the primal-dual witness construction approach, which is used for proving our main recovery result, and how we instantiate this framework under our statistical assumptions.  Section~\ref{sec:conclusion} provides a few brief concluding comments and discussion of some future directions.  Finally, the intermediate analytical results are relegated to the appendix.
}

\section{Main Theoretical Results}\label{sec:main-result}

{Our} main theoretical contribution {here comes in the form of a} new support recovery guarantee for the group Lasso estimator {under a random signal model}.  As alluded above, we assume measurements acquired according to the linear model \eqref{eq:obs-model}, and examine the performance of the group lasso estimator \eqref{GLasso} under the assumption that the unknown $\bbeta^*$ can be parsimoniously expressed in terms of a  given partition of its entries into blocks, as in \eqref{eq:grouping}.  

{
}

Our  recovery guarantee is expressed in terms of the inter-block and intra-block coherence parameters \cite{eldar2010block,bajwa2015conditioning} of the dictionary $\bX$ which are defined with respect to a given column-wise block partition of $\bX$.

\begin{definition}\label{def:blck-coh} For any dictionary $\bX=[\bX_{\cI_1} \bX_{\cI_2} \cdots \bX_{\cI_G}]$ with blocks $\bX_{\cI_g} \in \RR^{n\times d_g}$ {and whose columns all have unit Euclidean norm}, the \emph{inter-block coherence constant} $\mu_B(\bX)$ is defined as 
\begin{equation}\label{eq:block-coh}
\mu_B(\bX) := \max_{1\le g \neq g' \le G} \|\bX_{\cI_g}^T \bX_{\cI_{g'}}\|_{2\rightarrow 2},
\end{equation}
and the \emph{intra-block coherence parameter} $\mu_I(\bX)$ is defined as
\begin{equation}\label{eq:intra-block-coh}
\mu_I(\bX) := \max_{g\in[G]} \|\bX_{\cI_g}^T \bX_{\cI_g} - \bI_{d_g\times d_g}\|_{2\rightarrow 2}.
\end{equation}
\end{definition}

Notice that $\mu_B(\bX)$ measures similarity between the blocks of $\bX$ and reduces to the standard coherence parameter when the groups over the dictionary columns are singletons. {Further}, $\mu_I(\bX)$ measures the deviation of the blocks $\{\bX_{\cI_g}\}_{g\in [G]}$ from orthonormal blocks. From the computational perspective, both coherence parameters can be computed in polynomial time for a given column-wise partitioned dictionary (unlike other quantities such as restricted isometry constant, which are widely used in proving similar recovery guarantees; but can be NP-hard to compute \cite{bandeira2013certifying}). 

{
}

To conduct our analysis, we impose some mild statistical assumptions on the generation of the coefficient vector $\bbeta^*$. 
Specifically, similar to \cite{bajwa2015conditioning}, we assume the group-sparse vector $\bbeta^\ast \in \mathbb{R}^{p}$ in \eqref{eq:grouping} is randomly generated according to the assumptions outlined below:

\begin{itemize}
\item[$M_1)$\hspace{-0.4em}] 
The block support $\cG^*$ of $\bbeta^*$
comprises $s:=|\cG^*|$ non-zero blocks, whose indices are selected uniformly at random from all subsets of $[G]$ of size $s$. 
\item[$M_2)$\hspace{-0.4em}]
The non-zero entries of  $\bbeta^*$ are equally likely to be positive or negative: $\EE  [\text{sign}(\bbeta^*_{j})] = 0$ for $j\in [p].$
\item[$M_3)$\hspace{-0.4em}]
The non-zero blocks of  $\bbeta^*$ have statistically independent ``directions.'' Specifically, it is assumed  that
$$
\Pr\left(\bigcap_{g \in \cG^*} \frac{\bbeta^*_{\cI_g}}{\|\bbeta^*_{\cI_g}\|_F}\in \cA_{g}\right)=\prod_{g \in \cG_i^*}\Pr\left(\frac{\bbeta^*_{\cI_g}}{\|\bbeta^*_{\cI_g}\|_F}\in \cA_{g}\right),
$$ where {for each $g$}, $\mathcal{A}_g \subset \mathbb{S}^{d_{g}-1}$ with $\mathbb{S}^{d_{g}-1}$ representing the unit sphere in $\RR^{d_g}$.
\end{itemize}  
 

The first generative assumption  $M_1$ prescribes how the group-level support $\cG^*$ of $\bbeta^*$ should be generated. Having generated the support, the next two assumptions impose very mild requirements on the generation of non-zero coefficients in this model. In particular, the second assumption $M_2$ requires the non-zero coefficients to have zero median and  $M_3$ requires the non-zero blocks of $\bbeta^*$ to be independent of each other. Finally, we note that the above assumptions allow for arbitrary statistical correlations between the coefficients that belong to the \textit{same} non-zero block.

\subsection{Main Result}

{Under the modeling assumptions above, we obtain the following theorem, whose proof appears in Section \ref{sec:main-proof}.
 
\begin{thmi}\label{thm:main-thm}
Consider the linear measurement model \eqref{eq:obs-model} with $\bw \sim \cN(0,\sigma^2 \bI_{n\times n})$.  Assume that  
\begin{align*}
1) &\, \mu_I(\bX) \le c_0 \ \mbox{ {and} } \  \mu_B(\bX) \le \sqrt{\frac{d_{\min}}{d_{\max}^2}}\, \frac{c_1}{\log p},\\
2) &\, s = |\cG(\bbeta^*)| \le \min  \left\{  \frac{c_2\, G} {\|\bX\|^2_{2\rightarrow 2} \log p} , \frac{d_{\min}}{d_{\max}^2} \, \frac{c_2'\, \mu_B^{-2}(\bX)}{ \log p}\right\},\\ \vspace{3 cm}
3) &\, \forall g\in \cG(\bbeta^*):\\ &\hspace{-0.6em}\|\bbeta_{\cI_{g}}^*\|_2 \ge 10 \sigma (1+\epsilon) (\sqrt{d_{\cG}^*}+\sqrt{d_{g}})\, \max\left\{1,\sqrt{\frac{s}{d_{\max}\, \log p}}\right\}\\
4) &\, \lambda_g = 4  \sigma (1+\epsilon) \sqrt{d_g},\; \forall g\in[G],
\end{align*}
all hold for some positive constants $c_0$, $c_1 \le {0.001}$, 
\begin{equation}\label{eq:constants}
c_2 \le {\left[\sqrt{9+\frac{1}{2}\left(\frac{1}{4} - 3c_0 - 48 c_1\right)}-3\right]^2},
\end{equation}
$c_2'=\min\{c_2,{0.0001}\}$, and some 
$$
 \epsilon \ge \sqrt{\frac{(1+\mu_I(\bX))\, \log(p\,G)}{d_{\min}}}.
$$
{Then the following hold simultaneously, with probability at least $ 1-12\, p^{-2\log 2}:$}
\begin{itemize}
\item the  solution $\widehat{\bbeta}$ of \eqref{GLasso} is unique and has the same group-level support as $\bbeta^*$; that is, $\cG(\bbeta^*) = \cG(\widehat{\bbeta})$, and
\item 
 $\left\|\hat{\bbeta}_{\cI_{g}}-\bbeta^*_{\cI_{g}}\right\|_2 \le 5\sigma (1+\epsilon) \left(\sqrt{d_{g}}+\sqrt{d^*_{\cG}}\right)$, $\forall g\in\cG(\bbeta^*)$.
\end{itemize} 
\end{thmi}

According to the first condition, the support recovery guarantee relies on the well-conditioning of the dictionary $\bX$. We measure the well-conditioning in terms of block coherence constants $\mu_I(\bX)$ and $\mu_B(\bX)$ of the dictionary. The condition on $\mu_I(\bX)$ implies that the blocks of the dictionary are close to being orthonormal. When all the constituent blocks are of the same size, i.e. $d_{\min} = d_{\max} = d$, the condition on $\mu_B(\bX)$ implies that the worst-case dissimilarity between the blocks should scale as $\cO(\frac{1}{\sqrt{d}\cdot\log p})$. This is the same condition as the one required for the exact recovery guarantee of Theorem 2 in \cite{bajwa2015conditioning}. 
As we will later discuss in the context of the material anomaly detection framework, this first assumption will impose mild conditions on the problem parameters.

The second condition  specifies the requirement on the maximum number of allowable non-zero groups in the group-level support of $\bbeta^*$ that can be recovered. The condition provided here is not stringent since the block coherence parameter appears in the upper-bound in the form of $\mu_B^{-2}(\bX)$, which is a significant improvement over similar results, e.g. in  \cite{lv2011group, eldar2010block}, that require $|\cG^*|$ be upper bounded by a term that is $\cO(\mu_B^{-1}(\bX))$. As we will argue in the next section, in the case where the dictionary $\bX$ is the concatenation of the $N$-dimensional identity and discrete cosine transform (DCT) bases, and the dictionary blocks are solely defined over one of the two bases, this condition implies that the number of recoverable groups can be as large as $|\cG(\bbeta^*)| = \cO(N)$. 

As an another example (motivated by a similar discussion in \cite{bajwa2015conditioning}) assume the case of equal-sized groups, i.e. when $d_{\max}=d_{\min}= d$, and notice that $\|\bX\|_{2\rightarrow 2}^2 \ge p/n$ for any dictionary $\bX$ with normalized columns. Then, ${G}/(\|\bX\|^2_{2\rightarrow 2}\cdot \log p) =\cO(\frac{n}{d\cdot \log p})$. The inequality in this case holds for tight frames. Moreover, as shown by Theorem 4 of \cite{calderbank2015block}, there exist block dictionaries for which $\mu_B(\bX)\ge \sqrt{\frac{d}{n}}$, for which it follows that ${1}/(\mu_B^2(\bX)\cdot \log p)=\cO(\frac{n}{d\cdot \log p})$. These two facts together imply that the condition required by the theorem does not suffer from the square-root bottleneck.

The third assumption is on the strength of the non-zero groups, which requires their magnitudes to be above a certain threshold depending on the noise variance $\sigma.$ Notice that,  in contrast to \cite{liu2009estimation}, the strength condition stated here is non-asymptotic. Moreover, notice that setting $\epsilon$ to the smallest value allowed by the theorem statement would lower the threshold on the strength of the non-zero groups. In that case, the regularization constants $\lambda_g$ can be set to smaller values as according to the condition 4.  More discussions on this assumption, {and its implications in our motivating application} are provided in Section~\ref{sec:exp-results}.

By imposing the regularization constant $\lambda_g$ to scale with $\sigma \sqrt{d_g}$ we are, in a sense, making sure that, after performing group-level soft thresholding, the noise component that impacts the estimation of this block is thresholded. This can be seen more clearly  when we study the optimality conditions of problem \eqref{GLasso} in Lemma \ref{lem:optimal} (also see \cite{huang2010benefit} for similar regularization conditions).

When applied to the special case of the Lasso, where $d_{\min}=d_{\max} = 1$,  the following simplifications are implied: $\mu_I(\bX) = 0$, $\mu_B(\bX) = \mu(\bX)$, where $$\mu(\bX):=\max_{1\le i < j \le p} |X_i^T X_j|$$ is the standard coherence parameter of $\bX$  \cite{candes2009near}, $d_{\cG}^* = s$, $G = p$, and $\epsilon = \sqrt{2\log p}$. Consequently, the sufficient conditions of Theorem \ref{thm:main-thm} reduce to $\mu(\bX) \le c_1/\log p$, $$s \le \frac{c_2{''} p }{\|\bX\|_{2\rightarrow 2}^2\,\log p}$$ where $c_2''$ is a function of $c_2$ and $c_2'$, for every $i\in\cS(\bbeta^*)$
\begin{equation}\label{eq:amp-lasso}
|\bbeta_i^*| \ge 10 \sigma (1+\sqrt{2\log p})\, (\sqrt{s}+1)\,\max\left\{1,\sqrt{\frac{s}{\log p}}\right\}
\end{equation}
and finally $\lambda = 4\sigma (1+\sqrt{2\log p})$. In comparison with \cite{candes2009near}, which provides a coherence-based support recovery guarantee for Lasso by leveraging fixed designs and similar statistical assumptions as ours, the signal strength condition required in Eq. \eqref{eq:amp-lasso} is more restrictive, since it requires $\min_{i\in\cS} |\beta_i^*| =\Omega(s)$ for $s\ge \log p$. When discussing the proof of the theorem in Section \ref{subsection:pdw-analysis}, we will indicate the cause of this difference.  


Finally, our choices of the universal constants $c_0, c_1, c_2, c_2'$ are not optimized here. The relationship between these constants in Eq. \eqref{eq:constants} is reminiscent of those appearing in \cite{bajwa2015conditioning}. Improving these constants is left  for a future work.

\section{Application: Structural Anomaly Detection}\label{sec:prb-frml}

Structural health monitoring is critical to reliability and cost effective life-management of physical structures. {Recently}, a powerful {new} class of diagnostic methodologies has {emerged, leveraging} the availability of laser-based sensing systems~\cite{Sharma-et-al_Damage-Index_AIAA_2006,Michaels_Ruzzene_Michaels_Ultrasonics_2010}. {Through the use of a} \emph{Scanning Laser Doppler Vibrometer} (SLDV), it is possible to perform non-contact measurements at a {large number} of points {on a \emph{scanning grid} defined on the surface of an object under test}, thus {providing} full spatial reconstruction of {the material's surface} dynamic response (e.g., to an induced acoustic excitation).

Laser-based methods {facilitate diagnostic methods in} which the inference is performed directly on a data-rich, spatially reconstructed response.  {Central to} this view is the notion that, from a data standpoint, a wavefield is a data cube, slices of which represent snapshots (or frames) of the dynamic response at different {temporal} instants.  The task of {locating} anomalies in a physical medium, {then, can be recast as a problem of identifying atypical patterns in the observed data structures}.  {Such efforts have been among the} essential themes in machine learning and computer vision in recent years (see, e.g., \cite{Chandola09}).

\subsection{Approach}\label{subsec-approach}

In this work, we utilize and expand notions from the sparsity-based source separation literature \cite{donoho2001uncertainty,elad2005simultaneous,amelunxen2014living,foygel2014corrupted} and group Lasso inference to analyze the damage localization problem. The key observation underlying our approach is that SLDV measurements {of a material subjected to narrowband acoustic excitation}, acquired in the vicinity of the anomalous regions, exhibit different spatiotemporal behavior than do those acquired in the bulk of the material. We therefore attempt to decompose the {acquired wavefield data} into two components, one of which is a spatially-localized component arising near the defected areas while the other one is a generally smooth component in the pristine bulk of the structure.
This facilitates a \emph{baseline-free}, agnostic inference approach whereby the locations of the defects in a material may be accurately estimated without a priori characterization of (a pristine version of) the medium.  This feature distinguishes our method from the recent efforts in the context of Lamb wave-based structural health monitoring in \cite{levine2014block, golato2016multimodal} that assume the knowledge of the propagation model over the structure.

In order to separate the two structurally-distinct components of each measurement frame, we assume that 
each component can be efficiently expressed as a linear combination of elements from an appropriate {fixed dictionary or basis}. 
Since defects are  generally spatially-localized,  an appropriate dictionary for the defects is the identity matrix (i.e., the discrete Dirac basis), which comprises elements that are zero at every location except for one.  {Likewise,} the \emph{Discrete Cosine Transform} (DCT) basis is one suitable basis for the smooth component of the response from the undamaged regions. In this sense, our model is reminiscent of the {basis} pairs utilized in the initial works on Basis Pursuit \cite{donoho1989uncertainty, donoho2001uncertainty}.

Assume that one {vectorized} snapshot of wavefield measurements, captured at time instant $t\in[T]$, is denoted by the vector $\by(t) \in \RR^{N}$, where  $N$ denotes the total number of acquired measurements. 
  Moreover, assume that the matrix $\bY = \left[\by(1)\, \by(2) \cdots \by(T)\right]\in \RR^{{N}\times T}$ stores all the measurement vectors for time instants $1$ to $T$.  Further, let  $\bX_{(1)} \in \RR^{N\times p_1}$ and $\bX_{(2)}\in \RR^{N\times p_2}$ represent the dictionaries that appropriately represent the spatially-smooth and sparse components, respectively. We assume the following measurement model
\begin{align}\label{eq:measurement-time-t}
\bY = \bX_{(1)}\bB_{(1)}^{\ast} + \bX_{(2)} \bB_{(2)}^{\ast}+ \bW, 
\end{align}
where $\bB_{(1)}^*\in\RR^{{N}\times T}$ and $\bB_{(2)}^*\in \RR^{{N}\times T}$ denote the corresponding coefficient matrices and $\bW\in \RR^{{N}\times T}$ represents noise and model uncertainties.  Here, the first term $\bX_{(1)}\bB_{(1)}^{\ast} $ represents the smooth component of measurements generated by the pristine bulk of the medium and $\bX_{(2)}\bB_{(2)} ^{\ast} $ models the defect component. Given this, the problem of  anomaly detection reduces to finding the support of the defect component $\bX_{(2)} \bB_{(2)} ^{\ast}$ {(or simply $\bB_{(2)} ^{\ast}$ when $\bX_{(2)} = \bI_{N\times N} $)}.

To further facilitate the task of detecting anomalies, we notice that anomalies manifest themselves as \emph{spatially-contiguous} pixel blocks of the overall anomaly vector. Therefore, we propose to define spatial groupings over the domain of the defect component and make use of a spatial block-sparsity-promoting technique over the anomalous component of the measurement decomposition. Imposing the spatial block-sparsity condition is justified by the fact that the bulk of a medium is undamaged and therefore most of the spatial blocks of the anomalous component should be zero blocks. Given the measurement model \eqref{eq:measurement-time-t}, with $\bX_{(2)} = \bI_{N\times N}$, the spatial grouping assumption can be imposed by partitioning each column of $\bB^\ast_{(2)}$ into $G_2 := N/D$ groups of size $D$, where the entries within a group are adjacent pixels in the two-dimensional representation of the measurements.

In addition, since the effect of anomalies is usually persistent across multiple consecutive measurement frames (i.e., across time), we further extend the spatial grouping {to} a \emph{spatiotemporal} one.  Given \eqref{eq:measurement-time-t}, this can be easily done  by extending the column-by-column partition over the entries in $\bB^*_{(2)}$ across several consecutive columns (frames) and therefore forming $G_2$ sub-matrices of size $D\times T$ over $\bB^*_{(2)}$, where now the entries of a sub-matrix are spatiotemporally adjacent. 
On the other hand, a \emph{temporal} grouping can be applied to the entries of the coefficient matrix $\bB^*_{(1)}$ corresponding to the smooth component, with the idea that the same frequencies (i.e. the same columns of the DCT dictionary) should appear in the decomposition of consecutive frames. So, $\bB^*_{(1)}$ can be partitioned into $G_1:=N$ sub-matrices of dimensions $1\times T$.
  Given these assumptions, we propose to estimate the true coefficient matrices $\bB_{(1)} ^{\ast}$ and $\bB_{(2)} ^{\ast}$ by $\hat{\bB}_{(1)}$ and $\hat{\bB}_{(2)}$, which are solutions of the following optimization problem 
\begin{align}\label{eq:GLasso-applied}
&\min_{\bB_{(1)},\bB_{(2)}\in \RR^{N\times T}} \bigg\{\frac{1}{2} \left\| \bY -  \bX_{(1)}\bB_{(1)} - \bX_{(2)} \bB_{(2)}\right\|_{F}^2 + \nonumber \\
& \lambda_1 \sum_{g_1 \in [G_1]}\left\| (\bB^T_{(1)})_{\cI_{g_1}}\right\|_F+ \lambda_2 \sum_{g_2\in [G_2]} \left\|( \bB^T_{(2)})_{\cI_{g_2}} \right\|_F\bigg\},
\end{align}
where $\lambda_1$ and $\lambda_2$ are positive scalars, {and
$g_1$ and $g_2$ index the blocks} of $\bB_{(1)}$ and $\bB_{(2)}$, respectively, which are formed according to the above grouping techniques. The group level support of $\hat{\bB}_{(2)}$ corresponds to the detected anomalies.

\subsection{Main Results}\label{subsec-mainres} 

 
 \textcolor{black}{To apply the theoretical results developed in section \ref{sec:main-result}, we adopt a vectorized representation of the measurements in \eqref{eq:measurement-time-t}. Specifically, we choose $\by\in \RR^n$, with $n:=NT$, to denote the measurement vector acquired by stacking all the $T$ columns of $\bY$ in one vector. Upon vectorizing the entire measurement model \eqref{eq:measurement-time-t}, the new representation becomes
 \begin{equation}\label{eq:vector-representation}
\by = \tilde{\bX}_{(1)} \bbeta_{(1)}^* + \tilde{\bX}_{(2)} \bbeta_{(2)}^* + \bw,
\end{equation}
 where $\by = \text{vec}\left(\bY\right)\in \RR^n$, $\bbeta_{(1)}^* = \text{vec}(\bB_{(1)}^*)\in\RR^{{n}}$, $\bbeta_{(2)}^* = \text{vec}(\bB_{(2)}^*)\in \RR^{{n}}$, $\bw = \text{vec}(\bW)\in \RR^n$ are vectors, with the vectorization operator $\text{vec}(\cdot)$ stacking the columns of the argument matrix into a single-column vector, and $\tilde{\bX}_{(1)}$ and $\tilde{\bX}_{(2)}$ are Kronecker-structured dictionaries given as  
 $$\tilde{\bX}_{(i)} = \bI_{T\times T}\otimes \bX_{(i)}, \text{ for } i=1,2.$$ Notice that after the vectorization, the previously-discussed partitions over the entries of $\bB_{(1)}^*$ and $\bB_{(2)}^*$ result in non-canonical groups, which are either of size $T$ (for the groups over the smooth component) or of size $DT$ (for the groups over the second spatially-sparse component). Using vector notation, the problem \eqref{eq:GLasso-applied} can be recast as 
\begin{align}\label{eq:GLasso-applied-v2}
\min_{\bbeta_{(1)},\bbeta_{(2)}\in \RR^{n}} \bigg\{\frac{1}{2} \left\| \by -  \tilde{\bX}_{(1)}\bbeta_{(1)} - \tilde{\bX}_{(2)} \bbeta_{(2)}\right\|_{2}^2 \nonumber \\
+ \lambda_1 \sum_{g_1 \in [G_1]}\left\|(\bbeta_{(1)})_{\cI_{g_1}}\right\|_2+ \lambda_2 \sum_{g_2\in [G_2]} \left\|( \bbeta_{(2)})_{\cI_{g_2}} \right\|_2\bigg\}.
\end{align} 
We may write the model \eqref{eq:vector-representation} in terms of the overall dictionary $\bX := \left[\tilde{\bX}_{(1)} \,|\, \tilde{\bX}_{(2)}\right] \in \RR^{n\times p}$, with {$p:=2n$}, and the overall coefficient vector $(\bbeta^*)^T := [(\bbeta_{(1)}^*)^T \,|\, (\bbeta_{(2)}^*)^T] \in \RR^{p}$ as
\begin{equation}
\by = \bX \bbeta^{\ast} + \bw,
\end{equation}
which is the linear measurement model discussed earlier.}

{Next} we summarize the implications of our main theoretical result, Theorem \ref{thm:main-thm}, for the anomaly detection scenario described above. As the theorem states, under the statistical assumptions {$M_1$, $M_2$, and $M_3$}, and some extra conditions on the number of anomalies and their severity, exact detection of anomalous groups is possible.
\begin{cori}\label{cor:implications} 
Consider the linear measurement model \eqref{eq:measurement-time-t} with $\bX_{(1)}$ and $\bX_{(2)}$ specialized to the two-dimensional DCT and identity matrices of size $N\times N$, respectively, and the entries of $\bW$ drawn independently  from  $\cN(0,\sigma^2)$. Suppose  $\bB^*:=[\bB_{(1)}^\ast \,|\, \bB_{(2)}^*]\in \RR^{N\times 2T}$ has $s$ randomly-selected non-zero groups according to assumptions $M_1$, $M_2$, and $M_3$, with $\cG_1^*$ and $\cG_2^*$ denoting sets of indices of the true nonzero groups for the smooth and sparse components, repsectively. If
\begin{align*}
1) \ & \sqrt{N}\ge \frac{2\, \log(2NT)}{c_1}\sqrt{D^3\,T} \\
2) \ & s \le  \frac{c_{2} N}{TD^3 \log(2NT)}\\
3) \ & \tiny{\forall {g\in \cG_1^*}:\left\|(\bB_{(1)}^{*T})_{\cI_{g}} \right\|_F \ge 10 \sigma  \sqrt{T}\left(1+ \sqrt{s_1 + s_2 D} \right)} 
    \\& \qquad \qquad \qquad   \times(1+\epsilon) \max\left\{1, \sqrt{\frac{s}{TD\, \log(2NT)}}\right\} \\
4) \ & \tiny{\forall {g\in \cG_2^*}: \left\|(\bB_{(2)}^{*T})_{\cI_{g}} \right\|_F \ge 10 \sigma \sqrt{T} \left( \sqrt{D}+ \sqrt{s_1 + s_2 D} \right)} 
\\& \qquad \qquad \qquad  \times(1+\epsilon) \max\left\{1, \sqrt{\frac{s}{TD\, \log(2NT)}}\right\} \\ 
5) \ & \lambda_1 = 4\sigma(1+\epsilon)\sqrt{T} \textrm{ and } \lambda_2 = 4\sigma(1+\epsilon)\sqrt{DT}
\end{align*}
all hold for {$c_1 \le 0.001$}, $c_2 \le 0.0001$, and $$ \epsilon \ {\geq}\   \sqrt{\frac{2\log(2NT)}{T}},$$ {where $s_1$ and $s_2$ denote the number of nonzero groups selected in $\bB_{(1)}^\ast$ and $\bB_{(2)}^*$ respectively, then the group-level support of $\hat{\bB}$ will exactly match that of $\bB^\ast$ with probability at least $  1-12\, (2NT)^{-2\log 2}$.}
\end{cori}
\normalsize
The above {result}, whose proof is {provided} in Section \ref{sec:cor:implications},
is a direct consequence of Theorem \ref{thm:main-thm} of the previous section. 
The first condition simply suggests that the problem dimension needs to be sufficiently large for our results to be valid; this essentially ensures that the coherence conditions are satisfied for the specified basis pair. The second condition {provides an} upper bound {on} how many {anomalous} groups can be detected by the convex demixing procedure in \eqref{eq:GLasso-applied}. Interestingly, the number of recoverable non-zero groups is proportional to the total number of measurements here. The third and fourth conditions give lower bounds for the strength of the non-zero groups in order for them to be detectable {using the group Lasso approach}.  We explore these relationships numerically in the next section, where we test the ability of the group Lasso formulation \eqref{GLasso} in recovering the non-zero coefficients $\bbeta^*$ for dictionary-based representation of the measurements. 

\section{Numerical Experiments}\label{sec:exp-results}
The first set of experiments that are presented here  are carried out using synthetically generated {data adhering to our overall modeling assumptions}; {the second utilizes simulated data from our motivating structural health monitoring application, obtained via finite element simulation methods}. 

\subsection{Phase Transition Diagram}

{We begin by examining} the relationship between the group-sparsity level of the unknown coefficient vector and the strength of non-zero groups sufficient for recovery. The inspiration for this investigation comes from Conditions 3 and 4 of Corollary~\ref{cor:implications}, which {outline sufficient lower bounds} on $\|(\bB^{*T}_{(1)})_{\cI_g}\|_2$ and $\|(\bB^{*T}_{(2)})_{\cI_g}\|_2$ for exact support recovery. 

Operating under the measurement model assumptions introduced in {Section} \ref{sec:prb-frml}, we generate measurements  according to Equation \eqref{eq:measurement-time-t}. More specifically, we generate $T = 8$ frames of measurements, each of dimension $100\times 100$, therefore $N=10^4$ in \eqref{eq:measurement-time-t}. To generate each frame we choose $\bX_{(1)}$ to be the $N \times N$ 2D-DCT matrix,  and set $\bX_{(2)}$ to be the $N\times N$ identity matrix.
Once  $\bX_{(1)}$ and $\bX_{(2)}$ are selected, it remains to generate $\bB^*_{(1)} \in\RR^{N\times T}$, $\bB^*_{(2)} \in \RR^{N\times T}$ and $\bW \in \RR^{N\times T}$ in order to make the measurement vectors as according to \eqref{eq:measurement-time-t}.

 Inspired by the spatial contiguity assumption of anomalies, we assume each column of $\bB^*_{(2)}$, which corresponds to a vectorized $100\times 100$ image, is  partitioned into groups of size $D = d^2$, where each group corresponds to a $d\times d$ spatially-contiguous block in the original image representation of the column. Here we report the results for $d=2$ ($D= 4$). Also by the assumption of the temporal persistency of anomalies, we extend the grouping across all the frames resulting in the entries of $\bB_{(2)}^*$  be partitioned into groups of size $d^2\times T$. Doing so, the total number of blocks over the support of $\bB_{(2)}^*$ {becomes} $G_2 = (N/d)^2$. For the coefficient matrix $\bB^*_{(1)}$ corresponding to the spatially-smooth component,  we assume no spatial grouping structure over its columns; therefore each of its $G_1 = N = 10^4$ rows comprise a group.  Next, in order to give values to $\bB^* = [\bB^*_{(1)}\; \bB^*_{(2)}]$ we first choose  $s= s_1 + s_2$ out of the entire $G = G_1 +G_2$ blocks uniformly at random (for $s$ ranging from $1$ to $800$), set the selected entries to i.i.d. standard Gaussian values, and normalize each group to have magnitude $\alpha$. Finally, the noise matrix $\bW$ is set to have i.i.d. entries generated according to $\cN(0,1)$.  Thus, $\alpha$ can be thought as the parameter which defines the signal to noise ratio, and is varied from $0$ to $80$.  
 
For each choice of the {$(s,\alpha)$} pair, we generate $100$ different realizations and test the performance of the proposed algorithm in recovering the coefficients. The numerical algorithm that we have adopted for solving the corresponding optimization problem \eqref{eq:GLasso-applied} is {based on alternating minimization with respect to two coefficient matrices $\bB_{(1)}$ and $\bB_{(2)}$. A pseudocode sketch of the algorithm is detailed in our earlier work \cite{kadkhodaie2015locating}.  We note that the objective in \eqref{eq:GLasso-applied} is jointly-convex in $\bB_{(1)}$ and $\bB_{(2)}$. Applying the analysis  presented in \cite{kadkhodaie2014linear} it can be shown that the alternating minimization algorithm converges linearly to the global minima $(\bB_{(1)}, \bB_{(2)})$. We also provide a fully-documented {\tt MATLAB} software package called \emph{Damage Pursuit} to supplement this work; it is available for download at {\tt \url{http://damagepursuit.umn.edu}}}.


  Since by the specific grouping defined over the entries of $\bB_{(1)}^*$ and $\bB_{(2)}^*$ only two distinct group sizes exist (see Eq. \eqref{eq:GLasso-applied-v2}), the regularization parameters $\lambda_g$ for ${g\in[G]}$ are set to either $\lambda_1 = \frac{5}{\alpha} \sqrt{T}$ for all the groups defined over the support of $\bB_{(1)}^*$ or to $\lambda_2 = \frac{5}{\alpha} \sqrt{T\,d^2}$ for all the groups over the support of $\bB_{(2)}^*$.  The probability of success is then simply defined as the ratio of the number of  realizations for which the successful recovery of the group-level support of both $\bB_{(1)}^*$ and $\bB_{(2)}^*$  occurs to the total number of trails. To avoid errors due to numerical inaccuracies, we declare the groups of the recovered coefficient matrices as being non-zero if their norms exceed a precision constant $\epsilon_p =10^{-6}$ times the norms of their corresponding groups in the ground-truth coefficient matrices.

Fig. \ref{fig:FE-exp} (a) shows the phase transition diagram for the described set up. {As the number of active non-zero groups increases, one needs to increase the strength of the active groups to enable successful group-level support recovery, as expected. The shape of the curve shows agreement with our theoretical predictions.  Indeed, examining conditions 3 and 4 in Corollary~\ref{cor:implications}, we see that for small $s=s_1 + s_2$, the SNR above which the group Lasso succeeds is on the order of $\sqrt{s_1 + s_2D}$, while for larger values of $s$, the sufficient SNR condition is on the order of $\sqrt{s_1 + s_2D}\cdot \sqrt{s}$ (ignoring leading constants and other factors not depending on $s_1$ and $s_2$).  Now, because $D$ is small ($D=4$ here), we have that the small-$s$ regime should exhibit a sufficient SNR trend that functionally grows like $\sqrt{s}$, and the trend should be nearly linear in $s$.  This agrees, empirically, with the observed phase transition. Fig \ref{fig:FE-exp} (b) depicts a similar phenomenon from the same experimental data. Here, for every group-sparsity value, we find the signal strength value that  corresponds to the success probability nearest to 0.5, and plot the strength value versus the group-sparsity level on a log-log scale. The linear trend in this plot also suggests a polynomial relationship between the two quantities.}}

\subsection{Finite Element Simulations}
We also use synthetic wavefield measurements generated by finite element simulations to study the relationship between the number of defects, {their severity}, and the ability of the proposed group Lasso estimator in successful defect recovery. {For this}, we model an aluminum plate with dimensions $100$ cm $\times$ $50$ cm and thickness $5$ mm, which is probed by a flexural wavefield induced by an actuator located in the middle of the left edge of the domain. Localized anomalies are introduced by reducing the Young's modulus constant of the material of a  1.5 cm $\times$ 1.5 cm region to simulate a soft inclusion. The actuator is set to generate $N_c = 5$ bursts of a narrow-band sine wave at the frequency $f_c = 10^5$. We then record $100$  (two milliseconds apart) snapshots of the nodal displacements, over a grid with 160 $\times$ 80 nodes, and store them as columns of a measurements matrix $\bY$. Given the grid size and the number of frames, the measurement matrix $\bY$ {has} dimension ${N\times 100}$, where $N = 160\times 80 =12800$. 
 \begin{figure*}
 	\centering
 	\begin{tabular}{ccc}
	        \includegraphics[scale = 0.31]{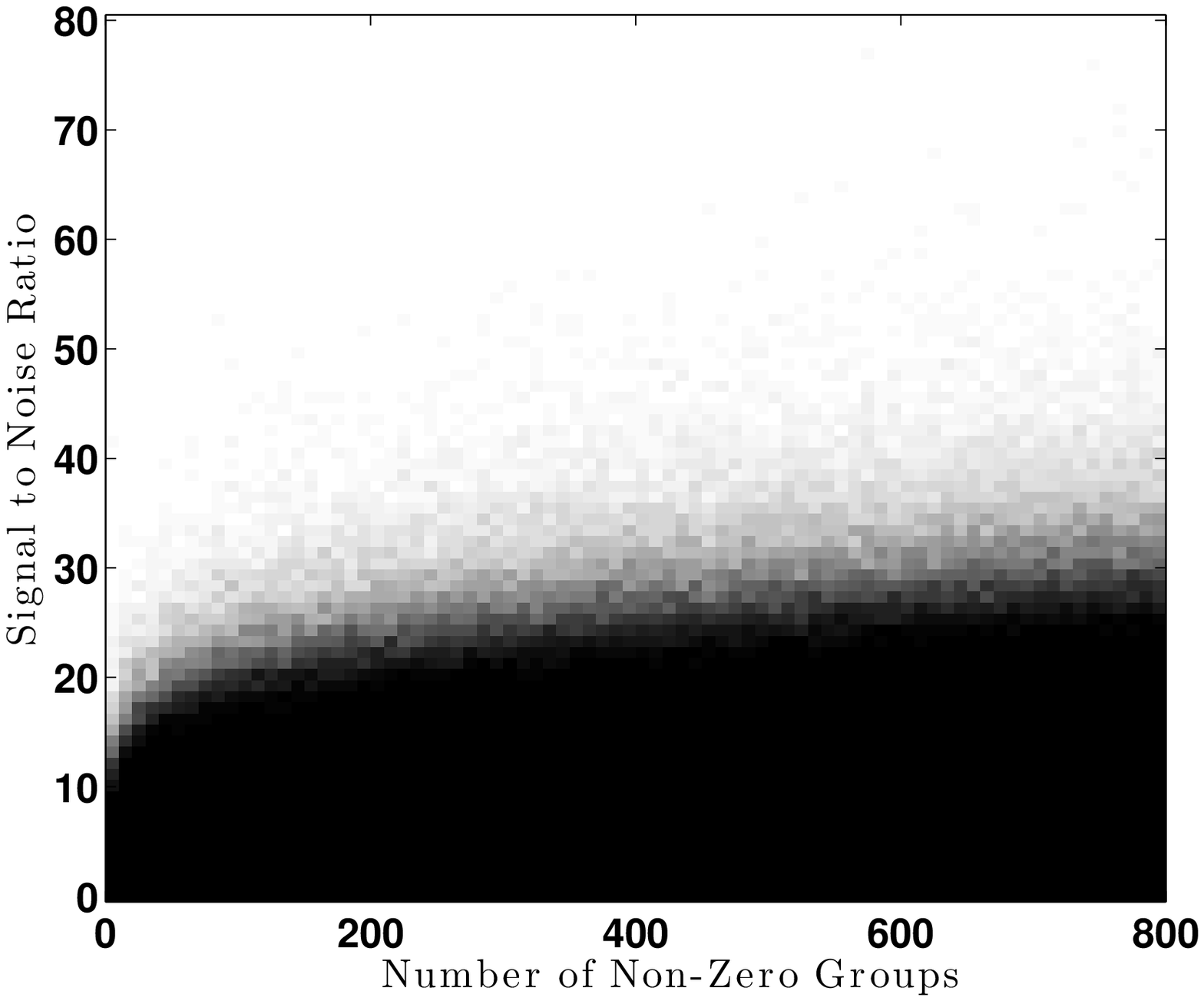} &
	        \includegraphics[scale = 0.19]{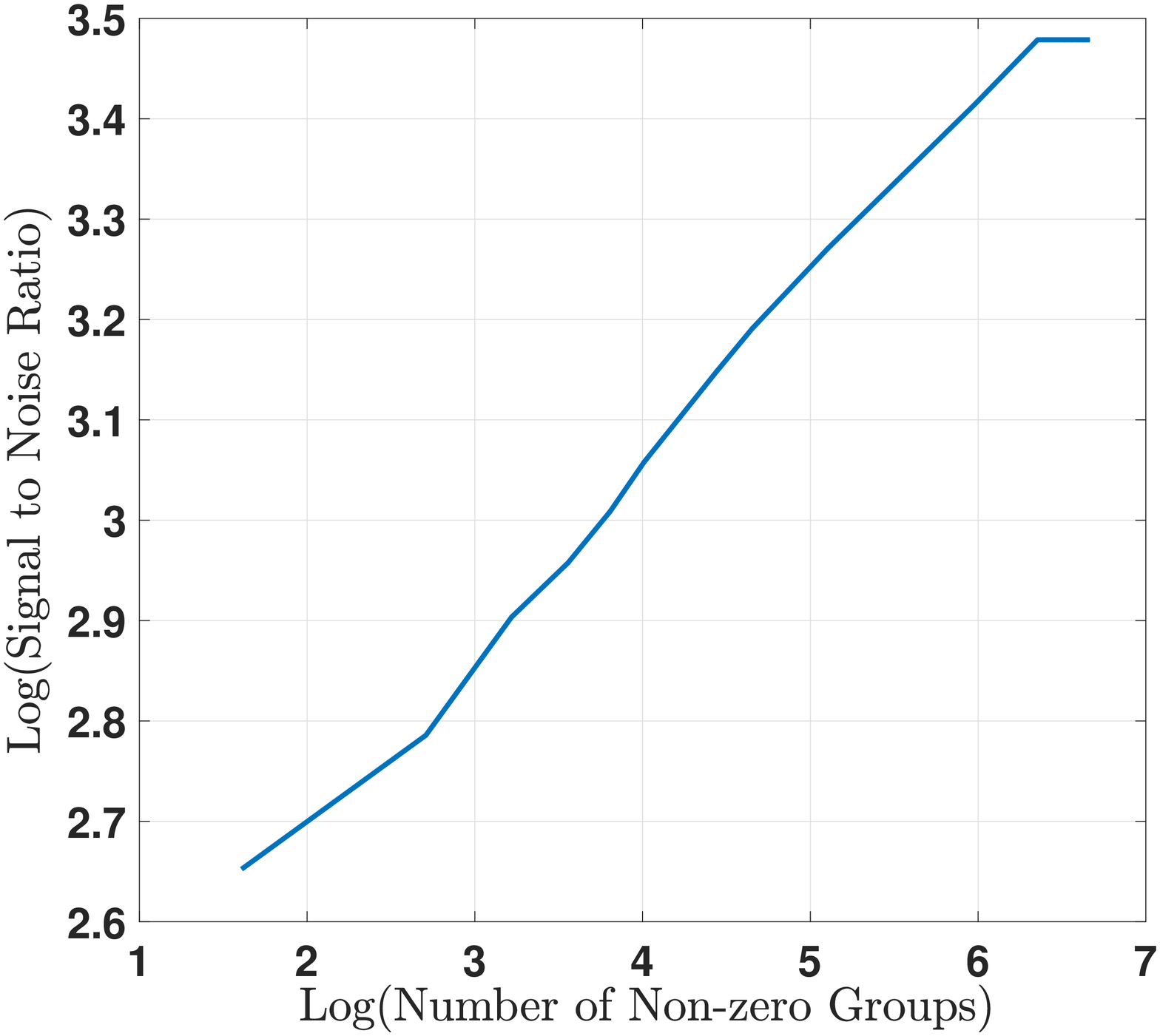}& 
 		\includegraphics[scale = 0.31]{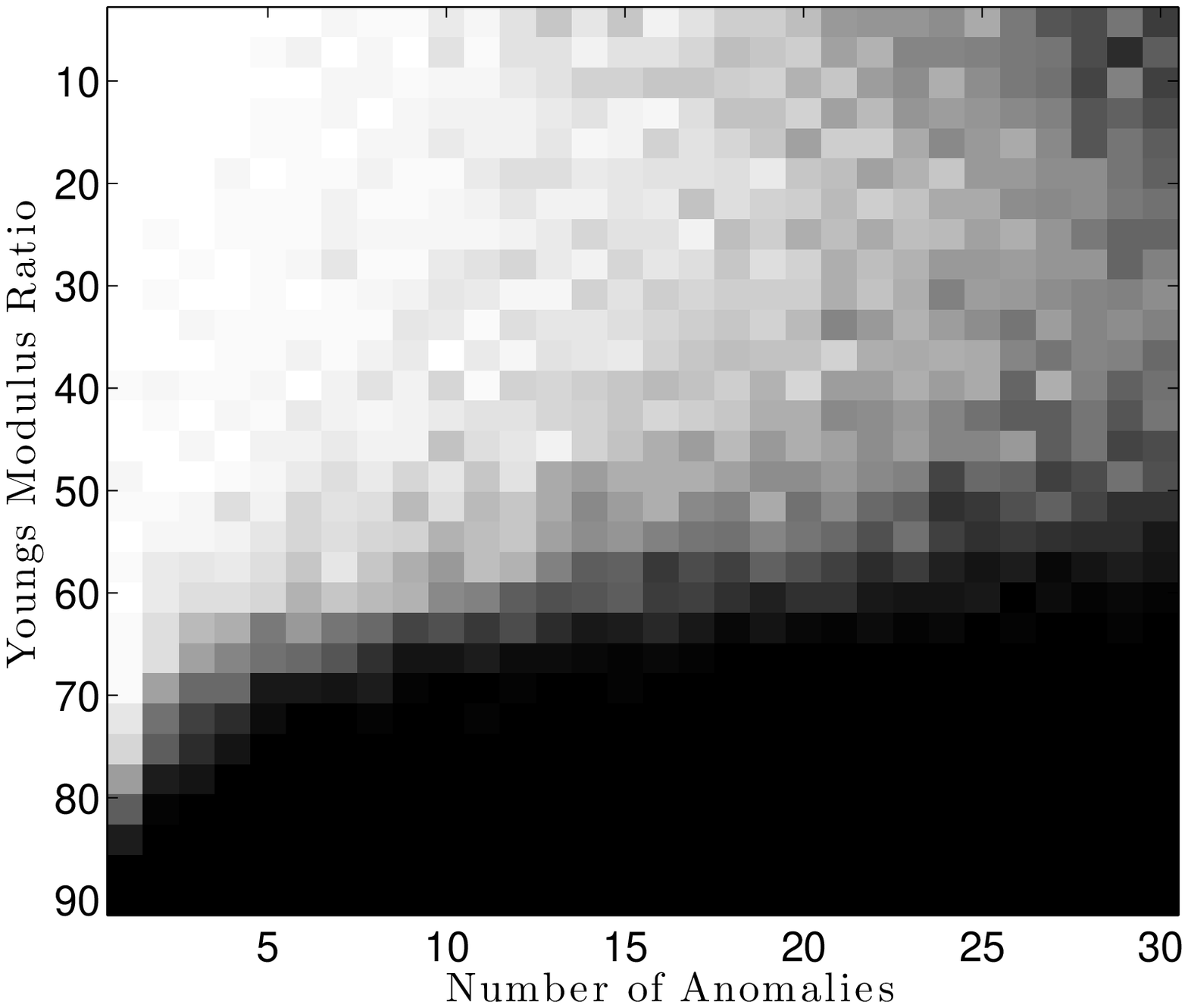}
		\\
 		(a) & (b) & (c)
 	\end{tabular}
 \caption{Panel (a) from left to right shows the phase transition diagram for the experiment with synthetically generated Gaussian data (white: success; black: failure). The vertical axis denotes the value  of signal to noise ratio varied through the scalar $\alpha$.  Panel (b) shows the transition boundary for the success probability of 0.5, on a log-log scale. (See text for discussion.) Panel (c) shows the resulting phase transition diagram from finite element experiments. The vertical axis denotes the ratio of the Young's modulus constant of defects to the bulk of the medium. The smaller this ratio is, the more severe the anomalies. 
} 
        \label{fig:FE-exp}
 \end{figure*}


Similar to the previous sub-section, we {aim} to generate a phase transition diagram for the successful recovery rate of our procedure, with the horizontal and vertical axes  indicating the number of defects and their severity level, respectively. This time, we vary the number of anomalies between $1$ and $30$, {and} place them at randomly selected locations over the  surface of the simulated structure. To change defects' severity {at those locations, we modulate} the Young's modulus constant of the bulk structure {by a} scalar parameter $\eta \in(0,1)$ {to obtain} the Young's modulus constant of {the} defected regions.   On the vertical axis of the phase transition diagram the defect severity is changed by raising $\eta$ to different integer powers $i$, where $i$ takes values between $1$ and $30$. As the integer power $i$ increases the defect severity increases as well, since the Young's modulus constant of defected regions become a smaller fraction of that corresponding to the healthy regions of the structure, which in turn makes defects more pronounced. 

In the current experiment we set $\eta=0.9$. We solve the  problem in \eqref{eq:GLasso-applied} for five consecutive frames, i.e. $T=5$, and adopt a partitioning of the defect component coefficient vectors into spatial groups of size four pixels. The regularization parameters were experimentally tuned to $\lambda_1= 0.005$ and $\lambda_2 = 0.12$ for the groups over the smooth and sparse components, respectively. We repeat the experiment 50 times for every specialization of the number of defects and their severity level. Fig \ref{fig:FE-exp} (c) shows the phase transition diagram for this experiment. Interestingly, the overall trend of the phase transition diagram resembles the diagram of the former sub-section. 
In fact, by increasing the mismatch between the Young's modulus constant of defects and the rest of the medium, local displacements at the place of anomalies increase. 
The displacements are effectively captured by the sparse coefficient matrix of our decomposition model and therefore contribute to stronger coefficient values in this matrix. 

Finally, we would like to note modifying the Young's modulus is but one principled approach to adjust the strength of an anomaly in a physical setting. Properly speaking, by adjusting this parameter, we are varying the contrast in elastic properties (acoustic mismatch). By extension, we can also model partial holes {via this approach} (see \cite{druce2017locating}), {but omit those evaluations here due to space constraints}.


\section{Proof of Theorem \ref{thm:main-thm}}\label{sec:main-proof}
In this section we present the proof of Theorem \ref{thm:main-thm}. For the purpose of clarity, the proof of the auxiliary lemmata that are required to show Theorem \ref{thm:main-thm} are relegated to the appendix.  
\subsection{Overview of Approach}
Our analysis utilizes a basic result for characterizing the optimal solutions of the group Lasso problem \eqref{GLasso}.  We state the result here as a lemma; its proof follows what are now fairly standard methods in convex analysis so we omit it here\footnote{A bit more specifically, we note that the proof of the lemma mirrors that of \cite[Lemma 1]{wainwright2009sharp}, with appropriate changes arising from the group Lasso regularizer. We also note that an analogous result appears, for example, in \cite{bach2008consistency, ravikumar2009sparse}, among other works.}. 
\begin{lemmai}\label{lem:optimal}
A vector $\check{\bbeta}$ solves problem \eqref{GLasso} if and only if 
\begin{equation}\label{KKT}
\bX_{\cI_g}^{T}\bX(\check{\bbeta}-\bbeta^*) -\bX_{\cI_g}^{T}\bw + \lambda_g \check{\bz}_{\cI_g} = \mathbf{0}, \ \ \forall \ g\in[G]
\end{equation} 
holds for some vector $\check{\bz}\in\RR^{p}$, whose elements satisfy
\begin{equation}\label{KKT:sub-grd}
\begin{array}{rl}
\check{\bz}_{\cI_g} = \check{\bbeta}_{\cI_g}/\|\check{\bbeta}_{\cI_g}\|_2, & \mbox{if }  \check{\bbeta}_{\cI_g} \neq \mathbf{0}\\
\|\check{\bz}_{\cI_g}\|_2 \leq 1, & \mbox{otherwise}
\end{array} .
\end{equation}
If $\|\check{\bz}_{\cI_g}\|_2 < 1$ for all $g\notin\cG(\check{\bbeta})$ then any optimal solution $\check{\bbeta}$ to \eqref{GLasso} satisfies $\check{\bbeta}_{\cI_g} = \mathbf{0}$ for all $g\notin\cG(\check{\bbeta})$; if, in addition, the matrix  $\bX^T_{\cS(\check{\bbeta})}\bX_{\cS(\check{\bbeta})}$ is invertible, then $\check{\bbeta}$ is the unique solution to \eqref{GLasso}.
\end{lemmai}
Note that the optimality condition \eqref{KKT} can be written in matrix form, as 
\begin{equation}\label{KKT_matrix}
\bX^{T}\bX(\check{\bbeta}-\bbeta^*) -\bX^T\bw + \bLambda \, \check{\bz} = \mathbf{0},
\end{equation}
where $\bLambda$ is the $p\times p$ diagonal matrix whose $j$-th diagonal entry $\Lambda_{j,j} = \lambda_{g(j)}$, where $g(j) = \{g\in[G]: j\in\cI_g\}$. In other words, the diagonal elements of $\bLambda$ are, for each index $j$, the regularization parameters associated with the group to which the corresponding element $\bbeta_j$ of $\bbeta$ belongs.  We will find this formulation convenient in the analysis that follows.

The ultimate goal of this section is to find conditions under which the group-level support of $\check{\bbeta}$ and $\bbeta^*$ are identical, i.e. $\cG(\check{\bbeta}) = \cG(\bbeta^*)$. Our proof follows the so-called \emph{Primal-Dual Witness} (PDW) technique utilized in \cite{wainwright2009sharp} for the analysis of the Lasso problem and also in \cite{ravikumar2009sparse, obozinski2011support} in related group Lasso problems.  In our setting, a primal-dual certificate pair $(\check{\bbeta},\check{\bz})$ is constructed according to the following steps:
\begin{enumerate}
\item We identify the solution of a \emph{restricted} group Lasso problem over the true ``group-level'' support $\cS_{\cG}(\bbeta^*)$. Specifically, we consider  $\check{\bbeta}_{\cS_{\cG}^*}\in\RR^{d^*_{\cG}}$ obtained {via}
\begin{equation}\label{rstc-glasso}
\check{\bbeta}_{\cS_{\cG}^*}=\arg\min_{\bbeta_{\cS_{\cG}^*}\in\RR^{d^*_{\cG}}} \frac{1}{2}||\by-\bX_{\cS_{\cG}^*}\bbeta_{\cS_{\cG}^*}||_2^2+\sum_{g\in \mathcal{{\cG^*}}}\lambda_g \|\bbeta_{\cI_g}\|_2.
\end{equation}
Note that if $\bX_{\cS_{\cG}^*}$ has full column-rank, there will be a unique vector $\check{\bbeta}_{\cS_{\cG}^*}$ that solves \eqref{rstc-glasso}.

\item We choose $\check{\bz}_{\cS_{\cG}^*}\in \RR^{d^*_{\cG}}$ to be the optimal dual solution of the restricted group Lasso problem \eqref{rstc-glasso} such that the primal-dual pair $(\check{\bbeta}_{\cS_{\cG}^*},\check{\bz}_{\cS_{\cG}^*})$ satisfies the optimality conditions of the restricted problem.

\item We set the ``off group-level support'' primal variable $\check{\bbeta}_{(\cS^*_{\cG})^c}$ to be zero. 

\item We solve for an ``off group-level support'' dual variable $\check{\bz}_{(\cS_{\cG}^*)^c}\in\RR^{p-d^*_{\cG}}$ which satisfies the optimality conditions for the full (unrestricted) group Lasso problem, as specified in \eqref{KKT} and \eqref{KKT:sub-grd}, and identify conditions under which this vector satisfies $\|\check{\bz}_{\cI_g}\|_2 < 1$ for all $g \notin \cG^*$.

\end{enumerate}
Overall, the PDW approach can be viewed, essentially, as a method for evaluating the feasibility of one particular candidate solution $\check{\bbeta}$ to the original group Lasso problem \eqref{GLasso}, constructed in a piece-wise manner.  The first two steps identify conditions that the elements of the candidate solution must adhere to on the true ``group-level'' support.  The strict dual feasibility condition ($\|\check{\bz}_{\cI_g}\|_2 < 1$ for all $g \notin \cG^*$) in Step 4 together with Step 3 ensure that no ``spurious'' nonzero groups are present in $\check{\bbeta}$.  In other words, the success of the PDW approach outlined above ensures that the primal-dual pair $(\check{\bbeta},\check{\bz})$ satisfies the optimality conditions of the general group Lasso problem \eqref{GLasso} as given by Lemma \ref{lem:optimal} and also meets the condition $\cG(\check{\bbeta}) \subseteq \cG^*$.

The last part of our analysis then relies on upper bounding the group-wise deviations between $\bbeta_{\cS_{\cG}^*}^*$ and $\check{\bbeta}_{\cS_{\cG}^*}$, from which we can identify conditions that the  nonzero groups of the true parameter vector $\bbeta^*$ must satisfy in order to ensure that no true signal groups are missed by the recovery procedure.  Specifically, suppose that the condition 
\begin{equation}\label{amp}
\|\bbeta^*_{\cI_g} - \check{\bbeta}_{\cI_g}\|_{2} < \|\bbeta_{\cI_{g}}^*\|_2 \ \ \mbox{ for all } g\in\cG^*
\end{equation}
holds true.  Then, it follows 
(essentially, by the triangle inequality) that $\check{\bbeta}_{\cI_g}\neq \mathbf{0}$, so that overall we have ${\bbeta}^*_{\cI_g}\neq \mathbf{0}$ implies $\check{\bbeta}_{\cI_g}\neq \mathbf{0}$. This is equivalent to $\cG^*\subseteq\cG(\check{\bbeta})$; overall, the success of the PDW method \emph{in addition to} a guarantee of the form \eqref{amp} will ensure that $\cG(\check{\bbeta})=\cG^*$. 

\subsection{Well-Conditioning of the Sub-Dictionary }
As we alluded when explaining the steps of the PDW approach, if $\bX_{\cS_{\cG}^*}$ has full column-rank, then the constructed $\check{\bbeta}$ will be the \emph{unique} optimal solution  of \eqref{GLasso};
see also Lemma 2 in \cite{obozinski2011support}. 
Throughout our analysis, we {condition on the event} that the  singular values of the block sub-dictionary $\bX_{\cS_{\cG}^*}$ lie within the interval $[\sqrt{1/2},\sqrt{3/2}\,]$. In other words, we  assume that the event
\begin{equation}\label{eqn:cond}
E_1:=\left\{\|\bX_{\cS_{\cG}^*}^T \bX_{\cS_{\cG}^*} - \bI_{d^*_{\cG}\times d^*_{\cG}}\|_{2\rightarrow 2} \leq \frac{1}{2}\right\}
\end{equation}
holds true. This event implies that the sub-dictionary $\bX_{\cS_{\cG}^*}$ is well-conditioned and full column-rank. 
Using the statistical {modeling} assumption ${M_1},$ we may obtain a probabilistic guarantee on the well-conditioning of $\bX_{\cS^*_{\cG}}$.  {This result, adapted from \cite{bajwa2015conditioning}, is stated below as a lemma}.
\begin{lemmai}[Adapted from \cite{bajwa2015conditioning}, Theorem 1]\label{lemma:well-conditioning}
Suppose that the $n\times p$ dictionary $\bX = \left[\bX_{\cI_1} \, \bX_{\cI_2} \cdots \bX_{\cI_G}\right]$
satisfies $\mu_I(\bX)\le c_0$ and $\mu_B(\bX)\le c_1/\log p$, with positive constants $c_0$ and $c_1$. Assume further that $\cG^*$ is a subset of size $s:=\left|\cG^*\right|$ of the set $[G]=\{1,2,\cdots,G\}$, drawn uniformly at random. Then, {provided}
\begin{equation}\label{eq:cond-for-well-cond}
s \le \min \left\{\frac{c_2}{\mu^2_B(\bX) \log p} , \frac{c_3\, G} {\|\bX\|^2_{2\rightarrow 2} \log p}\right\}
\end{equation}
for some positive constants $c_2$ and $c_3$ that only depend on $c_0$ and $c_1$, {we have that} the singular values of the sub-dictionary $\bX_{\cS_{\cG}^*}$ lie within the  interval $[\sqrt{1/2},\sqrt{3/2}]$, with probability at least $1-2p^{-4\log 2}$.
\end{lemmai} 
The above lemma is essentially identical to Theorem 1 of \cite{bajwa2015conditioning}, with the difference that in \eqref{eq:cond-for-well-cond} we have replaced {the so-called \emph{quadratic-mean block coherence} $\overline{\mu}_B(\bX)$ in \cite{bajwa2015conditioning} by $\mu_B(\bX)$.} This change yields a slightly more restrictive condition, since $\mu_B(\bX) \ge \overline{\mu}_B(\bX)$, but is sufficient for
our specific demixing problem.
  As a consequence of this lemma, it directly follows  that under the conditions {above}, $\|(\bX_{\cS_{\cG}^*}^T \bX_{\cS_{\cG}^*})^{-1}\|_{2\rightarrow 2} \leq 2$, {with high probability}. {Finally, we} note that $c_2$ and $c_3$ are selected here such that $(48 c_1 + 6\sqrt{2(c_2+c_3)} + 2 c_3 + 3 c_0) \le {1}/{4}$ holds true\footnote{This can be shown by using Eq. (5) in \cite{bajwa2015conditioning} and the discussion following that for bounding the expression appearing inside parentheses there.}. {This limits the allowable ranges of $c_0$ and $c_1$, as well.}

\subsection{Irrepresentablity of the Sub-Dictionary}
In addition to the well-conditioning event $E_1$, the PDW technique requires us to condition on the event that
$$
E_2 := \left\{\| \bX_{\cS_{\cG}^*}^T \bX_{(\cS_{\cG}^*)^c} \|_{B,1}
\le \gamma\right\}, 
$$
for the specific choice of 
\begin{equation}\label{eq:gamma-irrep-cond}
\gamma =  \frac{\lambda_{\min}}{\lambda_{\max}}\cdot\frac{c_4}{\sqrt{d_{\max}\cdot\log p}},
\end{equation} 
where $c_4$ is a positive constant (independent of problem parameters) that satisfies 
\begin{equation}\label{eq:c4}
c_4 \le \frac{1}{8\sqrt{2(1+4\log 2)}},
\end{equation}
 as required later in the proof. 
{When} this event holds,
we are ensured that blocks over the true group-level support $\cG^*$ are distinct enough from (or irrepresentable with) the remaining blocks. The following lemma, which is proved in the appendix, provides guarantees for this event.

\begin{lemmai}\label{lem:for-str-dual-cond}
Suppose the $n\times p$ dictionary $\bX$ is column-wise partitioned into $G$ blocks as $\bX = \left[\bX_{\cI_1} \, \bX_{\cI_2} \cdots \bX_{\cI_G}\right]$. Assume further that $\cG^*$ is a subset of the set $[G]$ of size $s=\left|\cG^*\right|$, drawn uniformly at random. Then, as long as
\begin{equation}\label{eq:for-str-dual-cond-2}
\mu_B(\bX) \le \frac{\gamma}{c_4}\cdot\min\left\{\frac{c_5}{\sqrt{\log p}}, \frac{c_6}{\sqrt{s}}\right\},
\end{equation}
where $\gamma $ is specified by \eqref{eq:gamma-irrep-cond} and $c_5$ and $c_6$ are small enough universal constants which satisfy $4\sqrt{2}\,c_5 + c_6 \le {c_4}/{2}$, we have
\begin{align}\label{eq:for-str-dual-cond}
\Pr(E_2)\ge 1-2\,p^{-4\log 2}.
\end{align}
\end{lemmai}

\subsection{Strict Dual Feasibility Condition}\label{subsection:pdw-analysis}
By Lemma \ref{lem:optimal},  $(\check{\bbeta},\check{\bz})$, with $\check{\bbeta}_{(\cS^*_{\cG})^c} = \mathbf{0}$, will be an optimal solution of the general group Lasso problem \eqref{GLasso} if and only if 

\begin{align}
&\bX_{\cS_{\cG}^*}^{T}\bX_{\cS_{\cG}^*}(\check{\bbeta}_{\cS_{\cG}^*}-\bbeta^*_{\cS_{\cG}^*}) -\bX_{\cS_{\cG}^*}^{T}\bw + \bLambda_{\cS_{\cG}^*} \check{\bz}_{\cS_{\cG}^*} = \mathbf{0},\label{eq1}\\
& \bX_{(\cS_{\cG}^*)^c}^{T}\bX_{\cS_{\cG}^*}(\check{\bbeta}_{\cS_{\cG}^*}-\bbeta^*_{\cS_{\cG}^*}) -\bX_{(\cS_{\cG}^*)^c}^{T}\bw + \bLambda_{(\cS_{\cG}^*)^c} \check{\bz}_{(\cS_{\cG}^*)^c} = \mathbf{0}\label{eq2},
\end{align}
where  $\bLambda_{\cS_{\cG}^*}$ and $\bLambda_{(\cS_{\cG}^*)^c}$  denote the sub-matrices of $\bLambda$ obtained by sampling rows and columns at the locations in ${\cS_{\cG}^*}$ and $({\cS_{\cG}^*})^c$, respectively, and $\check{\bz}$ satisfies the subgradient condition \eqref{KKT:sub-grd}.
Since $\bX_{\cS_{\cG}^*}^T \bX_{\cS_{\cG}^*}$ is invertible by the assumption that the event $E_1$ holds, we have that
\begin{equation}\label{eqn:implicit}
\bbeta^*_{\cS_{\cG}^*} - \check{\bbeta}_{\cS_{\cG}^*}   =  (\bX_{\cS_{\cG}^*}^T \bX_{\cS_{\cG}^*})^{-1}(\bLambda_{\cS_{\cG}^*} \check{\bz}_{\cS_{\cG}^*}-\bX_{\cS_{\cG}^*}^T \bw).
\end{equation}
Then, by step 4 of the PDW construction method, we take $\check{\bz}_{(\cS_{\cG}^*)^c}$ to be a vector that satisfies \eqref{eq2}. This gives that
\begin{equation*}
 \check{\bz}_{(\cS_{\cG}^*)^c} = \bLambda_{(\cS_{\cG}^*)^c}^{-1}\bX_{(\cS_{\cG}^*)^c}^{T}\bX_{\cS_{\cG}^*}(\bbeta^*_{\cS_{\cG}^*}-\check{\bbeta}_{\cS_{\cG}^*}) +
 \bLambda_{(\cS_{\cG}^*)^c}^{-1}\bX_{(\cS_{\cG}^*)^c}^{T}\bw,
\end{equation*}
and we now aim to establish the strict dual feasibility condition, that $\| \check{\bz}_{\cI_{g}}\|_2<1$ for all $g\notin\cG^*$. 
To that end, we note that for any fixed group index $g\notin\cG^*$ we have
\begin{align}
 \check{\bz}_{\cI_{g}}  
&=\frac{1}{\lambda_g}\bX_{\cI_{g}}^{T}\left[\bX_{\cS_{\cG}^*}(\bbeta^*_{\cS_{\cG}^*}-\check{\bbeta}_{\cS_{\cG}^*}) + \bw\right] \nonumber\\
&=\frac{1}{\lambda_g} \bX_{\cI_{g}}^{T}\left[ \bX_{\cS_{\cG}^*}(\bX_{\cS_{\cG}^*}^T \bX_{\cS_{\cG}^*})^{-1}(\bLambda_{\cS_{\cG}^*} \check{\bz}_{\cS_{\cG}^*}-\bX_{\cS_{\cG}^*}^T \bw) + \bw\right]\nonumber\\
&= \frac{1}{\lambda_g}\bX_{\cI_{g}}^{T}\left[ \bX_{\cS_{\cG}^*}(\bX_{\cS_{\cG}^*}^T \bX_{\cS_{\cG}^*})^{-1}\bLambda_{\cS_{\cG}^*} \check{\bz}_{\cS_{\cG}^*} +\Pi_{(\cS_{\cG}^*)^{\perp}}(\bw)\right],\nonumber
\end{align}
where the second equality follows from the incorporation of \eqref{eqn:implicit}, and the third one makes use of the definition $\Pi_{(\cS_{\cG}^*)^{\perp}}(\bw) := (\bI - \bX_{\cS_{\cG}^*}(\bX_{\cS_{\cG}^*}^T \bX_{\cS_{\cG}^*})^{-1}\bX_{\cS_{\cG}^*}^T) \bw$. 

Now, we exploit our statistical assumptions, i.e. that the ``direction'' vectors $\bbeta^*_{\cI_{g'}}/\|\bbeta^*_{\cI_{g'}}\|_2$ associated with every nonzero block of $\bbeta^*$ indexed by $g'\in\cG^*$ are random, and statistically independent.  To this aim, we express the vector $\check{\bz}_{\cS_{\cG}^*}$ (or more specifically, its individual blocks) in terms of the ``direction'' vectors associated with the corresponding nonzero blocks of the true vector $\bbeta^*_{\cS^*_{\cG}}$. The following lemma
 states that every block of $\check{\bz}_{\cS^*_{\cG}}$ is representable as the sum of the corresponding true direction vector and a bounded perturbation.

\begin{lemmai}\label{lem:dircond}
Suppose that the group-level support $\cG^*$ is fixed {and} that the event $E_1$ occurs. Defining $\bh_{g'} := \check{\bbeta}_{\cI_{g'}} - \bbeta^*_{\cI_{g'}}$ for every $g'\in\cG^*$, it follows that
\begin{equation}\label{eq:h-bound}
\|\bh_{g'}\|_2 \leq \|\bX^T_{\cI_{g'}}\bw\|_2 + \lambda_{g'} + \|\bX^T_{\cS_{\cG}^*}\bw\|_2 + \|\blambda_{\cG^*}\|_2,
\end{equation}
where $\blambda_{\cG^*}\in\RR^{d^*_{\cG}}$ is a vector whose entries are the elements $\{\lambda_{g'}\}_{{g'}\in\cG^*}$. Moreover, the blocks of the dual vector over the true support set $\cG^*$ can be expressed as
\begin{equation}\label{eq:directional-inconsistency}
\check{\bz}_{\cI_{g'}} = 
\frac{{\bbeta}^*_{\cI_{g'}}}{\|{\bbeta}^*_{\cI_{g'}}\|_2} + \bu_{g'},
\end{equation}
{and} if $\|\bh_{g'}\|_2 \le \frac{1}{2} \|\bbeta_{g'}^*\|_2$ for $g'\in \cG^*$, then  $\|\bu_{g'}\|_2 \le 4\|\bh_{g'} \|_2 / \|\bbeta^*_{\cI_{g'}} \|_2$.

\end{lemmai}

According to this lemma, which is shown in the appendix, for each $g'\in \cG^*$ {for which it holds that} $\|\check{\bbeta}_{\cI_{g'}} - \bbeta^*_{\cI_{g'}}\|_2 \le \frac{1}{2} \|\bbeta_{g'}^*\|_2$, we can write $\check{\bz}_{\cI_{g'}} = \left({\bbeta}^*_{\cI_{g'}}/{\|{\bbeta}^*_{\cI_{g'}}\|_2}\right) + \bu_{g'},$ where the norm of $\bu_{g'}$ can be controlled in terms of the norm of the difference  $\check{\bbeta}_{\cI_{g'}} - \bbeta^*_{\cI_{g'}}$. We can also express the condition \eqref{eq:directional-inconsistency} in the following compact form over the entire support $\cS_{\cG}^*$
$$
 \check{\bz}_{\cS_{\cG}^*} =\overline{\bbeta^*_{\cS_{\cG}^*}}+ \bu_{\cS_{\cG}^*},
$$
where $\overline{\bbeta^*_{\cS_{\cG}^*}}$ is obtained by concatenating the direction vectors ${{\bbeta}^*_{\cI_{g'}}}/{\|{\bbeta}^*_{\cI_{g'}}\|_2}$ for all $g'\in \cG^*$ and similarly $ \bu_{\cS_{\cG}^*}$ is the result of stacking all $ \{\bu_{g'}\}_{g'\in \cG^*}$.
 With this, we have overall that for each $g\notin\cG^*$, we can write 
\begin{eqnarray}
\nonumber \| \check{\bz}_{\cI_{g}}\|_2 &\leq& \frac{1}{\lambda_g}\left\|\bX_{\cI_{g}}^{T} \bX_{\cS_{\cG}^*}(\bX_{\cS_{\cG}^*}^T \bX_{\cS_{\cG}^*})^{-1} \bLambda_{\cS_{\cG}^*}\overline{\bbeta^*_{\cS_{\cG}^*}} \right\|_2  \\
\nonumber &+&  \frac{1}{\lambda_g}\left\|\bX_{\cI_{g}}^{T} \bX_{\cS_{\cG}^*}(\bX_{\cS_{\cG}^*}^T \bX_{\cS_{\cG}^*})^{-1} \bLambda_{\cS_{\cG}^*}\bu_{\cS_{\cG}^*}\right\|_2  \\
&+& \frac{1}{\lambda_g}\left\|\bX_{\cI_{g}}^{T}\Pi_{(\cS_{\cG}^*)^{\perp}}(\bw)\right\|_2.\label{eq:dual-bound}
\end{eqnarray}

{Now, by establishing} that the right-hand side is strictly less than $1$ for each $g\notin\cG^*$, {we ensure} no ``spurious'' groups will be identified by the group Lasso. This strategy {is central} to the proof of Theorem \ref{thm:main-thm}, which {employs} concentration arguments to control the terms in the above upper bound. 

{Before moving forward we note that, in the case where $d_g=1$ for all $g\in[G]$, i.e. for the standard Lasso, a stronger analysis is presented in \cite{candes2009near} that does not rely on defining the perturbation vectors $\bu_{g'}$. Interestingly, in that case the vectors $\check{\bz}_{\cI_{g'}}$ and ${{\bbeta}^*_{\cI_{g'}}}/{\|{\bbeta}^*_{\cI_{g'}}\|_2}$ become one-dimensional and reduce to the  signs of $\check{\bbeta}_{\cI_{g'}}$ and ${\bbeta}^*_{\cI_{g'}}$, respectively, that can be shown to be identical. Therefore it will readily follow that $\bu_{g'} =0$.  }

\subsection{Bounding the Terms in \eqref{eq:dual-bound}}\label{sec:prbt-anls}
{Now, conditioned} on the events $E_1$ and $E_2$, to prove the strict dual feasibility condition we will show that for any $g\notin \cG^*$, each of the terms appearing in the upper bound {in \eqref{eq:dual-bound}} can be further bounded (e.g. by the constant $1/4$) under the assumptions {$M_2$ and $M_3$} of our statistical model. To better organize the proof, we also define the three following probabilistic events, which correspond to the terms of the upper bound in \eqref{eq:dual-bound}: 

\begin{align}
E_3 &:=\left\{\left\|\bX_{\cI_{g}}^{T} \bX_{\cS_{\cG}^*}(\bX_{\cS_{\cG}^*}^T \bX_{\cS_{\cG}^*})^{-1} \bLambda_{\cS_{\cG}^*} \overline{\bbeta^*_{\cS_{\cG}^*}} \right\|_2 \le \frac{\lambda_g}{4}, \; \forall g\notin \cG^* \right\}\nonumber \\
E_4 &:=\left\{\left\|\bX_{\cI_{g}}^{T} \bX_{\cS_{\cG}^*}(\bX_{\cS_{\cG}^*}^T \bX_{\cS_{\cG}^*})^{-1} \bLambda_{\cS_{\cG}^*} \bu_{\cS_{\cG}^*} \right\|_2 \le \frac{\lambda_g}{4}, \; \forall g\notin \cG^* \right\}\nonumber \\
E_5 &:= \left\{ \|\bX_{\cI_{g}}^{T}\Pi_{(\cS_{\cG}^*)^{\perp}}(\bw)\|_2 \le \frac{\lambda_g}{4},\; \forall g\notin \cG^*\right\}.\nonumber
\end{align} 
Lemmata \ref{lem:berstein}, \ref{lem:hterm}, and \ref{lem:Gaussian-noise} {below} describe conditions under which these events {each} hold with high probability. {With these}, the {probabilistic guarantee of the} strict dual feasibility condition will naturally follow using a simple union bounding argument. The proofs of these lemmata are {in} the {appendix}.

\subsubsection{{Event} $E_3$}
The following lemma provides a condition under which the event $E_3$ 
{holds with high probability}. 

\begin{lemmai}\label{lem:berstein}
Suppose the group-level support $\cG^*$ is given such that the events $E_1$ and $E_2$ hold for the sub-dictionary $\bX_{\cS_{\cG}^*}$ of  the  dictionary $\bX\in \RR^{n\times p}$. Then assuming  $\bbeta_{\cS_{\cG}^*}^*$ is a random vector generated according to the statistical model assumptions {$M_2$ and $M_3$} described earlier we have that
\begin{eqnarray}\label{eq:Bernstein-bnd}
\lefteqn{\Pr\left(\bigcup_{g\notin\cG^*} \left\|\bX_{\cI_{g}}^{T} \bX_{\cS_{\cG}^*}(\bX_{\cS_{\cG}^*}^T \bX_{\cS_{\cG}^*})^{-1} \bLambda_{\cS_{\cG}^*}  \overline{\bbeta^*_{\cS_{\cG}^*}} \right\|_2    > \frac{\lambda_g}{4}\right)}\hspace{17em}&&\\
\nonumber &\leq& 2p^{-4\,\log 2}.
\end{eqnarray}
\end{lemmai}

\subsubsection{{Event} $E_4$}
Next, we derive conditions under which  the event $E_4$ 
holds {with high probability}. In order to show this, we leverage Lemma \ref{lem:dircond} to control the size of the $\{\bu_{g'}\}_{g'\in \cG^*}$ vectors and in turn the size of the $\{\bh_{g'}\}_{g'\in \cG^*}$ vectors. Since the upper bound in \eqref{eq:h-bound} for $\bh_{g'}$, ${g'\in \cG^*}$, is in terms of the noise-related terms $\|\bX^T_{\cS_{\cG}^*}\bw\|_2$ and $\|\bX^T_{\cI_{g'}}\bw\|_2$, we will start by providing probabilistic bounds on these quantities.

\begin{lemmai}\label{lem:applyHW}
Suppose the group-level support $\cG^*$ is fixed, {and} $\bw\sim\mathcal{N}(0,\sigma^2 \bI_{n\times n})$. There exists a universal constant $c_7\in(3,7)$ for which {the following} holds: for any $t \ge 1$ and $$\epsilon \geq \sqrt{\frac{(1+\mu_I(\bX))\log\left(p^t \,|\cG^*|\right)}{c_7\, d_{\rm min}}},$$ 
we have that
\begin{itemize}
\item $\|\bX^T_{\cS_{\cG}^*}\bw\|_2 \leq \sigma (1+\epsilon) \sqrt{d^*_{\cG}}$, and
\item $\bigcap_{g'\in\cG^*} \left\{\|\bX^T_{\cI_{g'}}\bw\|_2 \leq \sigma(1+\epsilon)\sqrt{d_{g'}}\right\}$
\end{itemize}
hold simultaneously {with probability at least} 
$1-2\, p^{-t}-2\,\exp\left(- c_7\epsilon^2 d^*_{\cG}/2\right)$.
\end{lemmai}

The proof of this lemma utilizes the Hanson-Wright inequality (see, e.g., Theorem 2.1 of \cite{rudelson2013hanson}).
Now, by using this lemma together with Lemma \ref{lem:dircond} we obtain the following result on the norm of the difference vectors $\bh_{g'} = \check{\bbeta}_{\cI_{g'}} - \bbeta^*_{\cI_{g'}}$ for $g' \in \cG^*$.
\begin{cori}\label{cori:h-bound}
Suppose the group-level support $\cG^*$ is given such that the event $E_1$ holds. Furthermore, assume that  $\bw\sim\mathcal{N}(0,\sigma^2 \bI_{n\times n})$. There exists a universal finite constant $c_7>0$ for which the following holds: for any $t \ge 1$ and $$\epsilon \geq \sqrt{\frac{(1+\mu_I(\bX))\cdot\log\left(p^t\,|\cG^*|\right)}{c_7\, d_{\rm min}}},$$ we have that
\begin{equation}
\|\bh_{g'}\|_2 \le  \sigma (1+\epsilon)  \left(\sqrt{{d^*_{\cG}}} +\sqrt{d_{g'}} \right) + \lambda_{g'} + \|\blambda_{\cG^*}\|_2 
\end{equation}
{holds simultaneously} for every $g' \in \cG^*$ with probability at least $1-2\, p^{-t}-2\exp\left(- c_7\epsilon^2 d^*_{\cG}/2\right)$.
\end{cori}
Leveraging the above Corollary, we are able to  bound the norm of the second term of the upper bound in \eqref{eq:dual-bound}.
\begin{lemmai}\label{lem:hterm}
Suppose the group-level support $\cG^*$ is given such that both events $E_1$ and $E_2 $ hold for the sub-dictionary $\bX_{\cS_{\cG}^*}$ of $\bX$. Furthermore, assume $\bw\sim\mathcal{N}(0,\sigma^2 \bI_{n\times n})$ and  that $\|\bbeta^*_{\cI_{g'}}\|_2 \geq t_2 \|\bh_{g'}\|_2$ holds for all $g'\in\cG^*$, for some value of $t_2$ satisfying 
\begin{equation}
t_2 \geq \max\left\{ 2, c_8 \sqrt{\frac{|\cG^*|}{d_{\max}\, \log p}}\right\},
\end{equation}
where $c_8$ is a universal constant which satisfies $c_8\ge4/\sqrt{2(1+4\log 2)}$. {Then,} we have that for all $g\notin\cG^*$
\begin{equation}
\frac{1}{\lambda_g}\left\|\bX_{\cI_{g}}^{T} \bX_{\cS_{\cG}^*}(\bX_{\cS_{\cG}^*}^T \bX_{\cS_{\cG}^*})^{-1} \bLambda_{\cS_{\cG}^*}\bu_{\cS_{\cG}^*}\right\|_2 \leq \frac{1}{4}.
\end{equation}
\end{lemmai}

Putting the result of Corollary \ref{cori:h-bound} together with the above lemma and also setting $c_8 = 2>4/\sqrt{2(1+4\log 2)}$, {we} immediately obtain the {following}.

\begin{cori}\label{cori:one}
Suppose the group-level support $\cG^*$ is given such that both events $E_1$ and $E_2 $ hold for the sub-dictionary $\bX_{\cS_{\cG}^*}$ of $\bX$. Furthermore, assume  $\bw\sim\mathcal{N}(0,\sigma^2 \bI_{n\times n})$, $\epsilon$ is set as in Theorem \ref{thm:main-thm}, and for all $g' \in \cG^*$
\begin{align*}\label{eq:beta-mag}
\|\bbeta_{\cI_{g'}}^*\|_2 &\ge\max\left\{2,2 \sqrt{\frac{|\cG^*|}{d_{\max}\cdot \log p}}\right\}\times \nonumber \\
&\left\{ \sigma (1+\epsilon)  \left(\sqrt{{d^*_{\cG}}} +\sqrt{d_{g'}} \right) + \lambda_{g'} + \|\blambda_{\cG^*}\|_2 \right\},
\end{align*}
{Then}
\begin{equation}
\left\|\bX_{\cI_{g}}^{T} \bX_{\cS_{\cG}^*}(\bX_{\cS_{\cG}^*}^T \bX_{\cS_{\cG}^*})^{-1} \bLambda_{\cS_{\cG}^*}\bu_{\cS_{\cG}^*}\right\|_2 \leq \frac{\lambda_g}{4},
\end{equation}
holds with probability at least $1-2\,p^{-t} -2\exp\left(- c_7\epsilon^2 d^*_{\cG}/2\right)$.
\end{cori}

\subsubsection{{Event} $E_5$} {Finally we} show that, with high probability, the noise-dependent term of the upper bound in \eqref{eq:dual-bound}, i.e. $\frac{1}{\lambda_g}\|\bX_{\cI_{g}}^{T}\Pi_{(\cS_{\cG}^*)^{\perp}}(\bw)\|_2$, is smaller than $1/4$ simultaneously for all $g\notin \cG^*$. See the {appendix} for the proof.

\begin{lemmai}\label{lem:Gaussian-noise}
Let $\bX$ be as above with $\cS_{\cG}^*$ fixed, and let $\bw\sim\cN(0,\sigma^2\bI_{n\times n})$. There exists a universal finite constant $c_7>0$ for which the following holds: for any $t \ge 1$ and $$\epsilon \geq    \sqrt{\frac{(1+\mu_I(\bX))\cdot\log\left(p^t \, (G-|\cG^*|)\right)}{c_7\,d_{\rm min}}}$$ if
$\lambda_g \geq 4\sigma(1+\epsilon) \sqrt{d_g}$, for all $g\notin\cG^*$,
then
\begin{equation}
 \Pr\left(\bigcup_{g\notin\cG^*} \left\{ \frac{1}{\lambda_g}\|\bX_{\cI_{g}}^{T}\Pi_{(\cS_{\cG}^*)^{\perp}}(\bw)\|_2 > \frac{1}{4}\right\}\right) \leq 2\,p^{-t}.
\end{equation}  

\end{lemmai}

\subsection{Completing the Proof of Theorem \ref{thm:main-thm}} 
Now we can put all the proof ingredients together to complete the overall argument. Let $E$ denote the event that the group-level support $\cG^*$ is exactly recovered via solving the group Lasso problem \eqref{GLasso}.  As explained in Section~\ref{sec:prbt-anls}, to ensure $E$ happens our approach is to first find conditions that guarantee $E_1$ and $E_2$ hold true; then conditioned on those two events, we impose extra assumptions to ensure $E_3,$ $E_4$, and $E_5$ occur as well. Using a union bound then implies the following upper bound\footnote{To show the inequality, notice that for two probabilistic events $A$ and $B$, we can write $A^c\cup B^c = A^c \cup (B^c\cap A)$. Setting $A = E_1 \cap E_2$ and $B = E_3\cap E_4 \cap E_5$ and using the fact that $\text{Pr}(A^c\cup B^c) \le \text{Pr}(A^c)+\text{Pr}(B^c\cap A)\le \text{Pr}(A^c)+\text{Pr}(B^c | A)$ {concludes the proof}.} 

\begin{align*}
\text{Pr}(E^c) &\le \text{Pr}(E_1^c) + \text{Pr}(E_2^c) + \text{Pr}\left(E_3^c \middle| E_1\cap E_2\right) \\
&+ \text{Pr}\left(E_4^c \middle| E_1\cap E_2\right) + \text{Pr}\left(E_5^c \middle| E_1\cap E_2\right).  
\end{align*}

The rest of the proof briefly reviews conditions under which the probability terms on the right-hand side of the above inequality are {appropriately} bounded. First, by Lemma \ref{lemma:well-conditioning} we know that if there exist positive constants $c_0$ and $c_1$ such that $\mu_I(\bX) \le c_0$, $\mu_B(\bX) \le {c_1}/{\log p}$, and 
\begin{equation}\label{eq:proof-s-first-const}
s \le \min \left\{\frac{c_2}{\mu^2_B(\bX) \log p} , \frac{c_3\, G} {\|\bX\|^2_{2\rightarrow 2} \log p}\right\},
\end{equation}
where $c_2$ and $c_3$ are such that 
\begin{equation}\label{eq:cnst-relation}
(48 c_1 + 6\sqrt{2(c_2+c_3)} + 2 c_3 + 3 c_0) \le \frac{1}{4},
\end{equation}
 then $\text{Pr}(E_1^c) \le 2p^{-4\log 2}$. Notice that the relationship in \eqref{eq:cnst-relation} requires $c_0$ and $c_1$ to be such that $48\,c_1 + 3\,c_0 \leq {1}/{4}$. Given this,  a valid choice for $c_2$ and $c_3$ that satisfies \eqref{eq:cnst-relation} is 
 $$
 c_2 = c_3 \le {\left[\sqrt{9+\frac{1}{2}\left(\frac{1}{4} - 3c_0 - 48 c_1\right)}-3\right]^2}.
 $$
 
 Second, utilizing Lemma \ref{lem:for-str-dual-cond}, with $\lambda_g = 4\sigma (1+\epsilon) \sqrt{d_g}$, $$\gamma :=  \frac{\lambda_{\min}}{\lambda_{\max}}\cdot \frac{c_4}{\sqrt{d_{\max}\cdot\log p}},$$ 
and $c_4$ {as in  the inequality in \eqref{eq:c4}},
it {follows} that as long as
 
\begin{equation}
 \mu_B(\bX) \le \sqrt{\frac{d_{\min}} {d_{\max}^2}} \frac{c_{5}}{\log p} \;\; \text{and}\;\; s \le \frac{d_{\min}}{d_{\max}^2}  \frac{c_6^2}{\mu_B^2(\bX)\,\log p},\label{eq:proof-s-second-const}
\end{equation}
 with constants $c_5$ and $c_6$ chosen such that $4\sqrt{2} c_5 + c_6 \le {c_4}/{2}$, then $\text{Pr}(E_2^c) \le 2p^{-4\log 2}$. In particular, 
 \begin{equation*}
 {c_5 = 0.001, \;  c_6 = 0.01}
 \end{equation*}
 are valid choices here.  To express the upper bounds in \eqref{eq:proof-s-first-const} and \eqref{eq:proof-s-second-const} on the maximum possible group-sparsity level $s$ more compactly, notice that since ${d_{\min}}/{d_{\max}^2} \le 1$, {we have that}
\begin{align*}
s&\le \frac{d_{\min}}{d_{\max}^2} \cdot \frac{\min\{c_6^2,c_2\}}{\mu_B^2(\bX) \log p}
\end{align*}
 together with $s\le{c_3\, G}\left/\left({\|\bX\|^2_{2\rightarrow 2} \cdot \log p}\right.\right)$ guarantees the requirements on $s$ are met. Similarly, $\sqrt{d_{\min}/{d^2_{\max}}}\le 1$  implies that imposing 
 $$\mu_B(\bX)\le  \sqrt{\frac{d_{\min}} {d_{\max}^2}}\cdot\frac{c_5}{\log p},$$ 
 will ensure the block coherence parameter meets $\mu_B(\bX) \le c_1/\log p$ for {$c_1 \leq 0.001$}.

Third, Lemma \ref{lem:berstein} implies that $\text{Pr}\left(E_3^c \middle| E_1\cap E_2\right)  \le 2p^{-4\log 2}$. {Fourth}, Corollary \ref{cori:one}, with $\lambda_g = 4\sigma (1+\epsilon)\sqrt{d_g}$ and $t=4\log2$, implies that as long as for 
\begin{equation}\label{low-bound:epsilon}
\epsilon \geq \sqrt{\frac{(1+\mu_I(\bX))\,\log\left(G\cdot p^{4\log 2}\right)}{c_7\,d_{\rm min}}}
\end{equation}
 we have 
\begin{align*}
\|\bbeta_{\cI_{g'}}^*\|_2 \ge  &10\,\sigma (1+\epsilon)  \left(\sqrt{{d^*_{\cG}}} +\sqrt{d_{g'}} \right)\times\\
&\max\left\{1, \sqrt{\frac{s}{d_{\max}\cdot \log p}}\right\} \; 
\end{align*}
for every $g'\in \cG^*$, then $\text{Pr}\left(E_4^c \middle| E_1\cap E_2\right) \le 2\,p^{-4\log 2} +2\exp\left(- c_7\epsilon^2 d^*_{\cG}/2\right)$.  Finally, by Lemma \ref{lem:Gaussian-noise}, we have that $\text{Pr}\left(E_5^c \middle| E_1\cap E_2\right) \le 2p^{-4\log 2}$ whenever $\lambda_g = 4\sigma(1+\epsilon) \sqrt{d_g}$ for all  $g\notin\cG^*$ . Hence, under {the} theorem conditions  we have
 \begin{align}
 \text{Pr}(E^c) \le 10p^{-4\log 2} +2\exp\left(- c_7\epsilon^2 d^*_{\cG}/2\right)
 \le 12\, p^{-2\log 2},\nonumber
 \end{align}
 where the last inequality follows from the lower bound on $\epsilon$, namely that $\exp(-c_7\epsilon^2 d_{\cG}^*/2) \le p^{-2\log 2}$. Finally, since $c_7 > 3$, the choice of $\epsilon$ in the theorem statement satisfies \eqref{low-bound:epsilon}.
 
 \section{Proof of Corollary \ref{cor:implications}}\label{sec:cor:implications} 

This is a direct consequence of Theorem \ref{thm:main-thm} for the anomaly detection framework studied in Section \ref{sec:prb-frml}. There we assumed $\bX = \left[\tilde{\bX}_{(1)} | \tilde{\bX}_{(2)} \right]$, where $\tilde{\bX}_{(1)} = \bI_{T\times T} \otimes \bX_{(1)}$ and $\tilde{\bX}_{(2)} = \bI_{T\times T} \otimes \bX_{(2)}$, with $\bX_{(1)}$ and $\bX_{(2)}$ specialized to two-dimensional DCT and identity matrices of size $N\times N$, respectively. Since in this setup $d_g$ is either $T$ (for the temporal groups defined over the support of the smooth component) or $DT$ (for the spatiotemporal groups defined over the support of the anomaly component), we set $\lambda_1 = 4\sigma(1+\epsilon)\sqrt{T}$ and $\lambda_2 = 4\sigma (1+\epsilon)\sqrt{DT}$ as in the statement of Theorem \ref{thm:main-thm}. Moreover, under the assumptions on the dictionary, we have $p=2NT$, $\|\bX\|_{2\rightarrow 2}^2 = 2$, the intra-block coherence parameter $\mu_{I}(\bX)$ will be zero and upper bounding $\mu_{B}(\bX)$ will amount to finding upper bounds on 
$$
\left\|\left(\tilde{\bX}_{(1)}\right)_{\cI_i}^T\left(\tilde{\bX}_{(2)} \right)_{\cI_j}\right\|_{2\rightarrow 2},
$$
 where $\left(\tilde{\bX}_{(1)}\right)_{\cI_i}$ and $\left(\tilde{\bX}_{(2)}\right)_{\cI_j}$ represent two column sub-matrices of $\tilde{\bX}_{(1)}$ and $\tilde{\bX}_{(2)}$ whose numbers of columns are given by the defined partition. More specifically, since the groups over the smooth component are temporal, we may write  $$\left(\tilde{\bX}_{(1)}\right)_{\cI_i} = \bI_{T\times T} \otimes \left(\bX_{(1)}\right)_{\cI_{i}} \in \RR^{NT\times T}$$ for the $T\times T$ identity matrix $\bI_{T\times T}$ and some column of $\bX_{(1)}$ denoted by $\left(\bX_{(1)}\right)_{\cI_{i}}$.  Also, since spatiotemporal groups are defined over the anomalous component, we may write $\left(\tilde{\bX}_{(2)}\right)_{\cI_j} = \bI_{T\times T} \otimes \left(\bX_{(2)}\right)_{\cI_j}$. Given these expressions for the sub-matrices of the two dictionaries, the associated inner products may be simplified as
 \begin{align*}
 \left(\tilde{\bX}_{(1)}\right)_{\cI_i}^T\left(\tilde{\bX}_{(2)} \right)_{\cI_j} &= \bI_{T\times T}\otimes  \left(\left( \bX_{(1)}\right)_{\cI_{i}}^T\left(\bX_{(2)} \right)_{\cI_j}\right),
 \end{align*}
{and it follows that} 
 $$\left\|\bI_{T\times T}  \otimes  \left(( \bX_{(1)})_{\cI_{i}}^T(\bX_{(2)})_{\cI_j}\right)\right\|_{2\rightarrow 2}=\left\|(\bX_{(1)})_{\cI_{i}}^T(\bX_{(2)})_{\cI_j}\right\|_{2}.$$
  
{Next, as} $\bX_{(1)}\in \RR^{N\times N}$ is a two-dimensional DCT matrix, the absolute value of {its largest entry} is no larger than $\sqrt{4/{N}}$; {see also \cite{candes2009near})}. Then since $\left(\bX_{(2)} \right)_{\cI_j}$ comprises $D$ columns of the identity matrix, the Euclidean norm on the right hand-side of the above expression will not exceed $\sqrt{{4D}/{N}}$. Therefore, the block coherence parameter satisfies $\mu_B(\bX) \le \sqrt{4D/N}$.  

The sufficient conditions stated in Corollary \ref{cor:implications} are then simplifications of the conditions in Theorem \ref{thm:main-thm}. In particular, that by imposing 
$
\sqrt{N}\ge \frac{2\, \log(2NT)}{c_1} \sqrt{D^3\,T},
$
 we are ensured $$\mu_B(\bX) \le \sqrt{\frac{d_{\min}}{d_{\max}^2}}\cdot \frac{c_{1}}{\log(2NT)}.$$ Furthermore,  the fact that $\mu_B(\bX) \le \sqrt{4D/N}$,  along with that ${d_{\max}^2}/{d_{\min}}=D^2\,T$, can be used to demonstrate
$$
 {\frac{d_{\min}}{d_{\max}^2}} \cdot \frac{c_2'\, \mu_B^{-2}(\bX)}{\log(2NT)} \ge \frac{c_2'}{4\log(2NT)}\cdot \frac{N}{TD^3}.
$$
 Then the condition  on the group-level sparsity in Theorem \ref{thm:main-thm} will be ensured by imposing
\begin{align*}
s=|\cG^*| &\le \frac{c_{2}' \,N}{4\,T D^3 \log(2NT)} \\
&= \min\left\{ \frac{c_2'\,N}{4\, TD^3 \log(2NT)} , \frac{c_2\,G}{2 \log(2NT)}\right\},
\end{align*}
since $G = N(1+{1}/{D})\ge N$, {$c_0=0$, and}
{
\begin{eqnarray*}
c_2 \leq 0.00028
\leq \left[\sqrt{9+\frac{1}{2}\left(\frac{1}{4} - 3c_0 - 48 c_1\right)}-3\right]^2
\end{eqnarray*}
so that $c_2'=0.0001=\min\{c_2,0.0001\}$.}

  \section{Discussion and Conclusions}\label{sec:conclusion}
In this paper we {examined} recovery of group-sparse signals from low-dimensional noisy linear measurements using the group Lasso  procedure, {motivated by a defect localization application in non-destructive evaluation}. {Our main theoretical result established new, practically {relevant}, non-asymptotic group-level support recovery guarantees in fixed dictionary settings.  Employing a mild statistical signal prior, our results improve upon existing results for such settings in terms of the number of nonzero groups that may be recovered, overcoming the well-known ``square root'' bottleneck from which deterministic coherence-based analyses are known to suffer.} We validated our analytical results via simulation on both synthetic data, and simulated data generated according to a realistic model for our motivating defect localization application.


\section{Appendix}\label{sec:appendix}
Here, we prove the lemmata that were utilized in the proof of the main Theorem.

\subsection{Proof of Lemma \ref{lem:for-str-dual-cond}}
We begin the proof by showing that for any $\gamma >0 $, we have 
\begin{align}\label{eq:for-str-dual-cond}
\Pr\bigg(&\bigg\| \bX_{\cS_{\cG}^*}^T \bX_{(\cS_{\cG}^*)^c} \bigg\|_{B,1} > \gamma \bigg)\nonumber\\
&\le 2 \,\left\{ \frac{\mu_B(\bX)}{\gamma} \cdot \left( 4{\sqrt{2\log p}}+ {\sqrt{s}} \right)\right\}^{4\log p}.
\end{align}

To show this we utilize Lemma A.5 in \cite{bajwa2015conditioning}, which implies
\begin{align*}
\text{Pr}\bigg(\bigg\| \bX_{\cS_{\cG}^*}^T \bX_{(\cS_{\cG}^*)^c}& \bigg\|_{B,1} > \gamma \bigg) \le 2 \gamma^{-q}\, \EE\left\|\bX_{\cS_{\cG}^*}^T \bX_{(\cS_{\cG}^*)^c} \right\|_{B,1} ^q \nonumber \\
&\le 2 \gamma^{-q}\, \EE\left\|\bX_{\cS_{\cG}^*}^T \bX\right\|_{B,1} ^q \nonumber \\
&\le 2 \gamma^{-q}\, \left( 2^{1.5} \sqrt{q}\, \mu_B(\bX) + \sqrt{s}\, \mu_B(\bX) \right)^q\nonumber 
\end{align*} 
where $q := 4\log p$, the first inequality is due to the Markov inequality and a Poissonization argument  ({a} similar argument is used in the proof of Theorems 1 and 2 in \cite{bajwa2015conditioning}), the second inequality is due to the fact that $\bX_{(\cS_{\cG}^*)^c}$ is a sub-dictionary of $\bX$, {and} the third inequality is by Lemma A.5 in \cite{bajwa2015conditioning} along with {the fact} that $\mu_B(\bX) \ge \overline{\mu}_B(\bX)$. Rearranging the terms then {completes} the proof of \eqref{eq:for-str-dual-cond}. 
{Now, setting} 
$$
\gamma =\frac{\lambda_{\min}}{\lambda_{\max}}\cdot \frac{c_4}{\sqrt{d_{\max}\cdot\log p}},
$$
where $c_4$ is an arbitrary positive constant,  will convert the upper bound of \eqref{eq:for-str-dual-cond} into
\begin{align*}
&\text{Pr}\bigg(\left\| \bX_{\cS_{\cG}^*}^T \bX_{(\cS_{\cG}^*)^c} \right\|_{B,1} > \gamma \bigg)\\
 &\le 2 \,\left( \frac{\mu_B(\bX)}{c_4}{{\frac{\lambda_{\max}}{\lambda_{\min}} \sqrt{d_{\max}\,\log p}}} \left( 4{\sqrt{2\log p}}+ {\sqrt{s}} \right)\right)^{4\log p}\nonumber\\
&\le 2 \left(4\sqrt{2}\, \frac{c_5}{c_4}+\frac{c_6}{c_4}\right)^{4\log p} \le 2\,p^{-4\log 2}
\end{align*}
where the second inequality is by imposing the following condition on $\mu_B(\bX)$:
\begin{equation}\label{eq:for-str-dual-cond-2}
\mu_B(\bX) \le \frac{\lambda_{\min}}{\lambda_{\max}}\cdot\frac{1}{\sqrt{d_{\max}\cdot \log p}}\cdot\min\left\{\frac{c_5}{\sqrt{\log p}}, \frac{c_6}{\sqrt{s}}\right\},
\end{equation}
 and the third one holds since $4\sqrt{2}\,c_5 + c_6 \le {c_4}/{2}$.

\subsection{Proof of Lemma \ref{lem:dircond}}

Using the relationship in \eqref{eqn:implicit} and defining $\bS_g\in \RR^{d_g\times d_\cG^*}$ as the selector matrix which selects indices corresponding to the block $g\in\cG^*$, we have that for each $g\in\cG^*$,
\begin{equation}
\check{\bbeta}_{\cI_g} = \bbeta^*_{\cI_g} + \bS_g (\bX_{\cS_{\cG}^*}^T \bX_{\cS_{\cG}^*})^{-1}(\bX_{\cS_{\cG}^*}^T \bw - \bLambda_{\cS_{\cG}^*} \check{\bz}_{\cS_{\cG}^*}).
\end{equation} 
Since the event $E_1$ is assumed to hold here, we  can write (by also using the Weyl's inequality) that $(\bX_{\cS_{\cG}^*}^T \bX_{\cS_{\cG}^*})^{-1} = \bI_{d^*_{\cG}\times d^*_{\cG}} + \bDelta$, where $\|\bDelta\|_{2\rightarrow 2} \le 1$. Then, note that
\begin{eqnarray*}
\lefteqn{\|\bh_g\|_2}&&\\
&=& \|\bS_g (\bX_{\cS_{\cG}^*}^T \bw - \bLambda_{\cS_{\cG}^*} \check{\bz}_{\cS_{\cG}^*}) + \bS_g \bDelta (\bX_{\cS_{\cG}^*}^T \bw - \bLambda_{\cS_{\cG}^*} \check{\bz}_{\cS_{\cG}^*})\|_2\\
&\leq& \|\bX_{\cI_g}^T \bw\|_2  + \|\lambda_g \check{\bz}_{\cI_g}\|_2 \\
&+& \|\bS_g\|_{2\rightarrow 2} \|\bDelta\|_{2\rightarrow 2} \big(\|\bX_{\cS_{\cG}^*}^T \bw\|_2 + \|\bLambda_{\cS_{\cG}^*} \check{\bz}_{\cS_{\cG}^*}\|_2\big).
\end{eqnarray*}
The {first} result follows from the facts that $\|\bDelta\|_{2\rightarrow 2}\leq 1$, and $\|\bS_g\|_{2\rightarrow 2} \leq 1$, and that
\begin{equation}
\|\bLambda_{\cS_{\cG}^*} \check{\bz}_{\cS_{\cG}^*}\|_2 = \bigg(\sum_{g\in\cG^*}\lambda_g^2 \|\check{\bz}_{\cI_g}\|_2^2\bigg)^{1/2}\leq \|\blambda_{\cG^*}\|_2,
\end{equation} 
where we have used the definition of $\blambda_{\cG^*}$, and the subgradient condition on each group of $\check{\bz}$.  

The second result follows from a similar argument as that given for Lemma 3 in \cite{obozinski2011support}.  To see this, in the statement of Lemma 3 in \cite{obozinski2011support} assume that the rows of the matrices $\Delta,  \hat{Z_S},$ and $\xi(B_S^*)$,  are set to $\bh_{g'}/\|\bbeta_{\cI_{g'}}\|_2$, $\check{\bz}_{\cI_{g'}}$, and ${\bbeta^*_{\cI_{g'}}}/{\|\bbeta^*_{\cI_{g'}}\|_2}$, respectively. We omit the proof to save space. 

\subsection{Proof of Lemma \ref{lem:berstein}}

The proof essentially follows the last step in the proof of Theorem 2 in \cite{bajwa2015conditioning}. First notice that {the event in} Eq. \eqref{eq:Bernstein-bnd}  is {equivalent} to the event that
\begin{align*}
\left\|\bLambda^{-1}_{\left(\cS_{\cG}^*\right)^c} \bX_{\left(\cS_{\cG}^*\right)^c}^{T} \bX_{\cS_{\cG}^*}(\bX_{\cS_{\cG}^*}^T \bX_{\cS_{\cG}^*})^{-1} \bLambda_{\cS_{\cG}^*} \overline{\bbeta^*_{\cS_{\cG}^*}} \right\|_{2,\infty}    > \frac{1}{4},
\end{align*}
 where for a block-wise partitioned vector $\ba := \left[\ba_{\cI_1}^T\, \ba_{\cI_2}^T\, \cdots \ba_{\cI_G}^T\right]^T$, $\|\ba\|_{2,\infty}$ is the maximum  Euclidean norm of its constituent blocks, i.e. $\|\ba\|_{2,\infty} := \max_{g\in [G]}\|\ba_{\cI_g}\|_2$. Furthermore, since $\|\ba\|_{2,\infty} \le \sqrt{d_{\max}} \|\ba\|_{\infty}$ with $d_{\max}$ denoting the maximum block size, it is sufficient to show that 
\begin{align*}
v&:=\left\|\bLambda^{-1}_{\left(\cS_{\cG}^*\right)^c} \bX_{\left(\cS_{\cG}^*\right)^c}^{T} \bX_{\cS_{\cG}^*}(\bX_{\cS_{\cG}^*}^T \bX_{\cS_{\cG}^*})^{-1} \bLambda_{\cS_{\cG}^*} \overline{\bbeta_{\cS_{\cG}^*}^*} \right\|_{\infty} \\
&\le \frac{1}{4\sqrt{d_{\max}}}.
\end{align*}
{holds with probability at least $1-2p^{-4\,\log 2}$}.

Letting $v_{g,j}:=\frac{1}{\lambda_g}\bx_{g,j}^T  \bX_{\cS_{\cG}^*}(\bX_{\cS_{\cG}^*}^T \bX_{\cS_{\cG}^*})^{-1} \bLambda_{\cS_{\cG}^*} \overline{\bbeta_{\cS_{\cG}^*}^*}  $, where $\bx_{g,j}$ denotes the $j$-th column in the block sub-dictionary $\bX_{\cI_g}\in \RR^{n\times d_g}$, with $j\in [d_g]$, we may write $v = \max_{g\notin \cG^*, \; j\in[d_g]}\left|v_{g,j}\right|$. Moreover, defining the vector $\bu_{g,j}:= \frac{1}{\lambda_g} (\bX_{\cS_{\cG}^*}^T \bX_{\cS_{\cG}^*})^{-1} \bX_{\cS_{\cG}^*}^T\, \bx_{g,j}$ for $g\notin \cG^*$ and $j\in[d_g]$, we can express each $v_{g,j}$ as {an} inner product of the form $$v_{g,j} = \bu_{g,j}^T\, \bLambda_{\cS_{\cG}^*} \,\overline{\bbeta_{\cS_{\cG}^*}^*}.$$ 

Notice that in this lemma we are proceeding under the condition that the selected block support $\cG^*$ is fixed,  and so {the} only random vector that appears on the right-hand side of the last expression is $\overline{\bbeta_{\cS_{\cG}^*}^*}$.  Now, by  the definition of  $\overline{\bbeta_{\cS_{\cG}^*}^*}$ and that $\bu_{g,j}$ is the concatenation of block vectors $\bu_{g,j,g'}\in \RR^{d_{g'}}$ (with $g'\in \cG^*$) corresponding to row-wise blocks in the partition of $(\bX_{\cS_{\cG}^*}^T \bX_{\cS_{\cG}^*})^{-1}$, we can express $v_{g,j}$ as
$$
v_{g,j} = \sum_{g'\in \cG^*} \lambda_{g'} \bu_{g,j,g'}^T \left(\frac{\bbeta^*_{\cI_{g'}}}{\|\bbeta^*_{\cI_{g'}}\|_2}\right).
$$
Since $v_{g,j}$ is now expressed in the form of the summation of random variables, its absolute value can be bounded by utilizing probabilistic concentration tools. To do so, first we apply {the} Cauchy-Schwartz inequality to every term in the summation to yield 
$$
\left|  \lambda_{g'} \bu_{g,j,g'}^T \left(\frac{\bbeta^*_{\cI_{g'}}}{\|\bbeta^*_{\cI_{g'}}\|_2}\right) \right| \le \lambda_{g'} \left\| \bu_{g,j,g'}\right\|_2,
$$
where we also employed the fact that ${\bbeta^*_{\cI_{g'}}}/{\|\bbeta^*_{\cI_{g'}}\|_2}$ is a unit-norm vector. Then since $\EE \left[{\bbeta^*_{\cI_{g'}}}/{\|\bbeta^*_{\cI_{g'}}\|_2}\right] = \zero$ for every $g'\in \cG^*$, Hoeffding's inequality {implies}
\begin{align*}
\text{Pr}\left( |v_{g,j}| \ge t\right) &\le 2 \exp\left( \frac{-t^2}{2\sum_{g'\in \cG^*} \lambda_{g'}^2 \|\bu_{g,j,g'}\|^2_{2} }   \right) \\
&= 2 \exp\left( \frac{-t^2}{2\| \bLambda_{\cS_{\cG}^*} \bu_{g,j}\|^2_{2}}  \right).
\end{align*}
{Now,} choosing $\kappa \ge \max_{g\notin \cG^*,\, j\in [d_g]} \| \bLambda_{\cS_{\cG}^*} \bw_{g,j} \|_2$ and applying a union bound we obtain 
$
\text{Pr} \left( v\ge t \right) \le 2 p \exp\left( {-t^2}/{2\kappa^2}\right). 
$
To find an appropriate choice for $\kappa$ that is explicitly in terms of our defining parameters, we explore upper bounds on $\bu_{g,j}$ as follows: 
$$
\|\bu_{g,j}\|_2 \le \frac{1}{\lambda_g} \left\| \left( \bX_{\cS_{\cG}^*}^T \bX_{\cS_{\cG}^*}\right)^{-1} \right\|_{2\rightarrow 2} \left\| \bX_{\cS_{\cG}^*}^T\, \bx_{g,j} \right \|_{2},
$$
where since $\bx_{g,j}$ is a column of the dictionary block $\bX_{\cI_g}$, it follows that 
\begin{align*}
\left\| \bX_{\cS_{\cG}^*}^T \bx_{g,j} \right \|_{2} &\le \left\| \bX_{\cS_{\cG}^*}^T \bX_{\cI_g} \right \|_{2\rightarrow 2} \le \max_{g\notin \cG^*}  \left\| \bX_{\cS_{\cG}^*}^T \bX_{\cI_g} \right \|_{2\rightarrow 2}\\
&\le \left\| \bX_{\cS_{\cG}^*} \bX_{\left(\cS_{\cG}^*\right)^c}   \right\|_{B,1}.
\end{align*}
Now, given that the selected sub-dictionary  is well-conditioned, i.e. $ \| ( \bX_{\cS_{\cG}^*}^T \bX_{\cS_{\cG}^*})^{-1} \|_{2\rightarrow 2} \le 2$, as guaranteed by $E_1$, and moreover that $\| \bX_{\cS_{\cG}^*} \bX_{(\cS_{\cG}^*)^c}\|_{B,1} \le \gamma$, as guaranteed by $E_2$, 
{we obtain that} $\|\bu_{g,j}\|_2 \le {2\gamma}/{\lambda_g}\le{2\gamma}/{\lambda_{\min}}$. Therefore an appropriate choice for $\kappa$ {is} $\kappa = 2\gamma({\lambda_{\max}}/{\lambda_{\min}}) $ (also utilizing the fact that $\|\bLambda_{\cS_{\cG}^*} \bu_{g,j}\|_2\le \lambda_{\max} \|\bu_{g,j}\|_2$). Now, setting $t={1}/({4\sqrt{d_{\max}}})$ and $$\gamma =  \frac{\lambda_{\min}}{\lambda_{\max}}\cdot\frac{c_4}{\sqrt{d_{\max}\cdot\log p}}$$ implies
\begin{align*}
\text{Pr} \left( v \ge \frac{1}{4\sqrt{d_{\max}}} \right) &\le 2p\cdot \exp\left(\frac{-1}{32\, \kappa^2\, d_{\max}}\right) \nonumber\\
&= 2 p\cdot \exp\left( \frac{-1}{128\, d_{\max} \, (\frac{\lambda_{\max}}{\lambda_{\min}})^2 \gamma^2  } \right)\\
&= 2 p^{\left(1-\frac{1}{128\, c_4^2}\right)} 
\end{align*}
{Thus}, assuming $c_4$ satisfies $1-\frac{1}{128\,c_4^2}\le -4\log 2$, {we have that} the last expression on the right hand-side  is less than  $2p^{-4 \log 2}$, which completes the proof.

\subsection{Proof of Lemma \ref{lem:applyHW}}

We establish that the events  $\left\{\|\bX^T_{\cS_{\cG}^*}\bw\|_2 \leq \sigma (1+\epsilon) \sqrt{d^*_{\cG}}\right\}$
and $ \left\{\|\bX^T_{\cI_{g'}}\bw\|_2 \leq \sigma(1+\epsilon)\sqrt{d_{g'}}\,,\; \forall g'\in \cG^*\right\}$ hold with the specified probability using the \emph{Hanson-Wright Inequality} \cite{rudelson2013hanson}, which states that 
for a fixed matrix $\bA$, and vector $\bx$  whose elements are iid $\cN(0,1)$ random variables (which are thus subgaussian), there exists a finite constant $c_7>0$ such that for any $\tau>0$, 
\begin{equation}
\Pr\left(\left| \|\bA\bx\|_2 - \|\bA\|_F\right| > \tau \right) \leq 2\exp\left(-\frac{c_7\tau^2}{\|\bA\|_{2\rightarrow 2}^2}\right).
\end{equation}
 

Now, fix any $g'\in\cG^*$ and note that
\begin{align*}
\Pr\left(\|\bX^T_{\cI_{g'}}\bw\|_2\right. &> \left. \sigma(1+\epsilon)\sqrt{d_{g'}}\right)\\
&\leq\Pr\left(\left| \|\bX^T_{\cI_{g'}}\bw\|_2 - \sigma\sqrt{d_{g'}} \right| > \epsilon \sigma \sqrt{d_{g'}}\right)\\
& \leq 2\exp\left(-\frac{c_7\, \epsilon^2 d_{g'}}{1+\mu_I(\bX)}\right),
\end{align*}
where the second inequality follows directly from {the} Hanson-Wright inequality ({specifically,} setting $\bx = \bw/\sigma$ and $\bA = \sigma \bX^T_{\cI_{g'}}$, and noting that $\|\bA\|_F = \sigma \sqrt{d_{g'}}$ and $\|\bA\|_{2\rightarrow 2} \le \sigma \sqrt{1+\mu_I(\bX)}$). Next, note that
\begin{align*}
\Pr\left(\|\bX^T_{\cS_\cG^*}\bw\|_2\right. &> \left.\sigma(1+\epsilon)\sqrt{d^*_{\cG}}\right)\\
&\leq \Pr\left(\left| \|\bX^T_{\cS_{\cG}^*}\bw\|_2 - \sigma\sqrt{d^*_{\cG}} \right| > \epsilon \sigma \sqrt{d^*_{\cG}}\right)\\
&\leq 2\exp\left(-\frac{2c_7\,\epsilon^2 d^*_{\cG}}{3}\right).
\end{align*}
Here, the second inequality follows again from {the} Hanson-Wright inequality, setting $\bx = \bw/\sigma$, and $\bA = \sigma \bX^T_{\cS_\cG^*}$, and noting that $\|\bA\|_F = \sigma \sqrt{d^*_{\cG}}$ (since each row of $\bA$ is unit-norm) and $\|\bA\|_{2\rightarrow 2} \leq \sigma \sqrt{3/2}$, which follows from event $E_1$.  

Thus, {by a union bound}, both of the stated claims hold, except in an event of probability no larger than $$2\exp\left(-2c_7\,\epsilon^2 d^*_{\cG}/3\right) + 2\sum_{g'\in\cG^*}\exp\left(-c_7\, \epsilon^2 d_{g'}/(1+\mu_I(\bX))\right),$$ which itself is upper-bounded by $$2\exp\left(-\frac{c_7\,\epsilon^2 d^*_{\cG}}{2}\right) + 2|\cG^*|\exp\left(-\frac{c_7 \epsilon^2 d_{\rm min}}{1+\mu_I(\bX)}\right),$$
where $d_{\rm min} := \min_{g\in[G]} d_g$. Finally, note that whenever 
\begin{equation}
\epsilon \geq \sqrt{\frac{(1+\mu_I(\bX))\cdot\log\left(p^t\,|\cG^*|\right)}{c_7\, d_{\rm min}}}
\end{equation}
for any $t\ge 1$, we have $$2|\cG^*|\exp\left(-\frac{c_7 \epsilon^2 d_{\rm min}}{1+\mu_I(\bX)}\right) \leq 2\,p^{-t},$$ and the result follows.

\subsection{Proof of Lemma \ref{lem:hterm}}

The sub-multiplicativity  of the spectral norm  obtains
\begin{align*}
&\frac{1}{\lambda_g}\big\|\bX_{\cI_{g}}^{T} \bX_{\cS_{\cG}^*}(\bX_{\cS_{\cG}^*}^T \bX_{\cS_{\cG}^*})^{-1} \bLambda_{\cS_{\cG}^*}\bu_{\cS_{\cG}^*}\big\|_2 \\
&\leq \frac{1}{\lambda_g}\left\|\bX_{\cI_{g}}^{T} \bX_{\cS_{\cG}^*}\right\|_{2\rightarrow 2}\, \left\|(\bX_{\cS_{\cG}^*}^T \bX_{\cS_{\cG}^*})^{-1}\right\|_{2\rightarrow 2}\, \left\|\bLambda_{\cS_{\cG}^*}\bu_{\cS_{\cG}^*}\right\|_2\nonumber \\
&\le \frac{2 \gamma}{\lambda_g}  \left\|\bLambda_{\cS_{\cG}^*}\bu_{\cS_{\cG}^*}\right\|_2 \le 2\gamma\,\frac{\lambda_{\max}}{\lambda_{\min}}\left\|\bu_{\cS_{\cG}^*}\right\|_2\nonumber
\end{align*}
where the second inequality follows since we assume  $E_1$ and $E_2$ hold true (therefore $\|(\bX_{\cS_{\cG}^*}^T \bX_{\cS_{\cG}^*})^{-1}\|_{2\rightarrow 2} \le 2$ and $\|\bX_{\cI_{g}}^{T} \bX_{\cS_{\cG}^*}\|_{2\rightarrow 2} \le \|\bX_{\cS_{\cG}^*}^T \bX_{(\cS_{\cG}^*)^c}\|_{B,1}\le\gamma$) and the third inequality follows by the fact that $\|\bLambda_{\cS_{\cG}^*}\|_{2\rightarrow 2} = \lambda_{\max}$ (and therefore $\|\bLambda_{\cS_{\cG}^*}\bu_{\cS_{\cG}^*}\|_2 \le \lambda_{\max} \|\bu_{\cS_{\cG}^*}\|_2$). In addition, note that by assuming  $\|\bbeta^*_{\cI_{g'}}\|_2 \geq t_2 \|\bh_{g'}\|_2 \ge 2\,\|\bh_{g'}\|_2$ for all $g'\in\cG^*$, Lemma \ref{lem:dircond} implies 
\begin{equation}
\|\bu_{g'}\|_2 \le 4\frac{\left\|\bh_{g'}\right\|_2}{\|\bbeta^*_{\cI_{g'}}\|_2} \leq \frac{4}{t_2}
\end{equation}
for all $g'\in\cG^*$ and therefore $\|\bu_{\cS_{\cG}^*}\|_2 \le {4\sqrt{|\cG^*|}}/{t_2}$.
Combining all of these results we obtain that 
\begin{align*}
\frac{1}{\lambda_g}\left\|\bX_{\cI_{g}}^{T} \bX_{\cS_{\cG}^*} (\bX_{\cS_{\cG}^*}^T \bX_{\cS_{\cG}^*})^{-1} \bLambda_{\cS_{\cG}^*}\bu_{\cS_{\cG}^*}\right\|_2 \le \frac{\lambda_{\max}}{\lambda_{\min}} \cdot \frac{8\gamma\sqrt{|\cG^*|}}{t_2}
\end{align*}
Therefore, assuming the event $E_2$ holds for the choice of $$\gamma = \frac{c_4}{\left(\frac{\lambda_{\max}}{\lambda_{\min}}\right) \sqrt{d_{\max}\cdot\log p}},$$ where $c_4 \le 1/8\sqrt{2(1+4\log 2)}$ is a finite positive constant  as appeared in the proof of Lemma  \ref{lem:berstein}, will ensure that 
$$
\frac{1}{\lambda_g}\left\|\bX_{\cI_{g}}^{T} \bX_{\cS_{\cG}^*} (\bX_{\cS_{\cG}^*}^T \bX_{\cS_{\cG}^*})^{-1} \bLambda_{\cS_{\cG}^*}\bu_{\cS_{\cG}^*}\right\|_2 \le \frac{8c_4}{t_2}\cdot \sqrt{\frac{{|\cG^*|}}{{d_{\max}\cdot\log p}}}.
$$ 
Then choosing $t_2 \ge c_8\sqrt{|\cG^*|/d_{\max}\log p}$ as specified by the statement of the lemma (with $c_8 := 32\,c_4$) {completes} the proof.

\subsection{Proof of Lemma \ref{lem:Gaussian-noise}}

Fix any $g\notin\cG^*$.  Note that for any $\tau>0$,
\begin{align*}
\Pr\left(\right.&\| \left. \bX_{\cI_{g}}^{T}\Pi_{(\cS_{\cG}^*)^{\perp}}\bw\|_2 > \sigma \sqrt{d_g} + \tau\right)\\ &\leq 
\Pr\left(\|\bX_{\cI_{g}}^{T}\Pi_{(\cS_{\cG}^*)^{\perp}}\bw\|_2 > \sigma\|\bX_{\cI_{g}}^{T}\Pi_{(\cS_{\cG}^*)^{\perp}}\|_F + \tau \right)\\
&\leq \Pr\left(\left|\|\bX_{\cI_{g}}^{T}\Pi_{(\cS_{\cG}^*)^{\perp}}\bw\|_2 - \sigma\|\bX_{\cI_{g}}^{T}\Pi_{(\cS_{\cG}^*)^{\perp}}\|_F\right| > \tau \right),
\end{align*}
where the first inequality follows from the fact that $\|\bX_{\cI_{g}}^{T}\Pi_{(\cS_{\cG}^*)^{\perp}}\|_F \leq \sqrt{d_g}$ (which is easy to verify by considering $\|\Pi_{(\cS_{\cG}^*)^{\perp}}^{T}\bX_{\cI_{g}}\|^2_F$, arranging the sum that arises in the definition of the squared Frobenius norm into a sum of sums over columns of $\bX_{\cI_{g}}$, and applying standard matrix inequalities along with the fact that $\|\Pi_{(\cS_{\cG}^*)^{\perp}}\|_{2\rightarrow 2} = 1$). 

Now, the final upper bound above is of the form controllable by the Hanson-Wright Inequality.  Specifically, setting $\bx = \bw/\sigma$, and $\bA = \sigma \bX_{\cI_{g}}^{T}\Pi_{(\cS_{\cG}^*)^{\perp}}$, and using the fact that $\|\bX_{\cI_{g}}^{T}\Pi_{(\cS_{\cG}^*)^{\perp}}\|_{2\rightarrow 2}\leq \sigma\sqrt{1+\mu_I(\bX)}$ (which is easy to verify using the sub-multiplicativity of the spectral norm), we obtain overall that for the universal finite constant $c_7>0$, and the specific choice $\tau = \epsilon \sigma \sqrt{d_g}$,
\begin{equation*}
\Pr\left(\|\bX_{\cI_{g}}^{T}\Pi_{(\cS_{\cG}^*)^{\perp}}\bw\|_2 > \sigma (1+\epsilon) \sqrt{d_g} \right) \leq 
2 \exp\left(-c_7\epsilon^2 d_g\right).
\end{equation*}
Thus, it follows that
\begin{align*}
\Pr\bigg(&\bigcup_{g\notin\cG^*} \left\{ \|\bX_{\cI_{g}}^{T}\Pi_{(\cS_{\cG}^*)^{\perp}}\bw\|_2 > \sigma (1+\epsilon) \sqrt{d_g}\right\} \bigg) \\
&\leq 2 \sum_{g\notin\cG^*} \exp\left(-c_7\epsilon^2 d_g\right)\nonumber\leq 2 (G-|\cG^*|) \exp\left(-c_7\epsilon^2 d_{\rm min}\right).\nonumber
\end{align*}
 Next, note that whenever 
$$
\epsilon \geq \sqrt{\frac{\log\left(p^t \, (G-|\cG^*|)\right)}{c_7\,d_{\rm min}}}
$$
the last term is no larger than $2\,p^{-t}$. Finally, note that the stated result holds if 
$
\lambda_g \geq 4\sigma (1+\epsilon) \sqrt{d_g} \ \ \mbox{ for all } g\notin\cG^*.
$

\bibliographystyle{IEEEbib}
\bibliography{grouprefs}

\begin{thebibliography}{10}

\bibitem{donoho2006compressed}
D.~L. Donoho,
\newblock ``Compressed sensing,''
\newblock {\em IEEE Transactions on Information Theory}, vol. 52, no. 4, pp.
  1289--1306, 2006.

\bibitem{candes2009exact}
E.~J. Cand{\`e}s and B.~Recht,
\newblock ``Exact matrix completion via convex optimization,''
\newblock {\em Foundations of Computational Mathematics}, vol. 9, 2009.

\bibitem{negahban2009unified}
S.~Negahban, B.~Yu, M.~J. Wainwright, and P.~K. Ravikumar,
\newblock ``A unified framework for high-dimensional analysis of m-estimators
  with decomposable regularizers,''
\newblock in {\em Adv. Neural Info. Proc. Sys.}, 2009.

\bibitem{chandrasekaran2012convex}
V.~Chandrasekaran, B.~Recht, P.~A. Parrilo, and A.~S. Willsky,
\newblock ``The convex geometry of linear inverse problems,''
\newblock {\em Foundations of Computational Mathematics}, vol. 12, no. 6, pp.
  805--849, 2012.

\bibitem{tropp2008conditioning}
J.~A. Tropp,
\newblock ``On the conditioning of random subdictionaries,''
\newblock {\em Applied and Computational Harmonic Analysis}, vol. 25, no. 1,
  pp. 1--24, 2008.

\bibitem{candes2009near}
E.~J. Cand{\`e}s and Y.~Plan,
\newblock ``Near-ideal model selection by $\ell_1$ minimization,''
\newblock {\em The Annals of Statistics}, vol. 37, no. 5A, 2009.

\bibitem{tibshirani1996regression}
R.~Tibshirani,
\newblock ``Regression shrinkage and selection via the {L}asso,''
\newblock {\em Journal of the Royal Statistical Society. Series B
  (Methodological)}, pp. 267--288, 1996.

\bibitem{yuan2006model}
M.~Yuan and Y.~Lin,
\newblock ``Model selection and estimation in regression with grouped
  variables,''
\newblock {\em Journal of the Royal Statistical Society: Series B (Statistical
  Methodology)}, vol. 68, no. 1, pp. 49--67, 2006.

\bibitem{bach2008consistency}
F.~R. Bach,
\newblock ``Consistency of the group {L}asso and multiple kernel learning,''
\newblock {\em The Journal of Machine Learning Research}, vol. 9, 2008.

\bibitem{tropp2006algorithms2}
J.~A. Tropp,
\newblock ``Algorithms for simultaneous sparse approximation. {P}art {II}:
  {C}onvex relaxation,''
\newblock {\em Signal Processing}, vol. 86, no. 3, 2006.

\bibitem{nardi2008asymptotic}
Y.~Nardi and A.~Rinaldo,
\newblock ``On the asymptotic properties of the group {L}asso estimator for
  linear models,''
\newblock {\em Electronic Journal of Statistics}, vol. 2, pp. 605--633, 2008.

\bibitem{meier2008group}
L.~Meier, S.~Van~De Geer, and P.~B{\"u}hlmann,
\newblock ``The group {L}asso for logistic regression,''
\newblock {\em Journal of the Royal Statistical Society: Series B (Statistical
  Methodology)}, vol. 70, no. 1, pp. 53--71, 2008.

\bibitem{liu2009estimation}
H.~Liu and J.~Zhang,
\newblock ``Estimation consistency of the group {L}asso and its applications,''
\newblock in {\em International Conference on Artificial Intelligence and
  Statistics}, 2009, pp. 376--383.

\bibitem{huang2010benefit}
J.~Huang and T.~Zhang,
\newblock ``The benefit of group sparsity,''
\newblock {\em The Annals of Statistics}, vol. 38, no. 4, pp. 1978--2004, 2010.

\bibitem{obozinski2011support}
G.~Obozinski, M.~J. Wainwright, and M.~I. Jordan,
\newblock ``Support union recovery in high-dimensional multivariate
  regression,''
\newblock {\em The Annals of Statistics}, pp. 1--47, 2011.

\bibitem{kolar2011union}
M.~Kolar, J.~Lafferty, and L.~Wasserman,
\newblock ``Union support recovery in multi-task learning,''
\newblock {\em Journal of Machine Learning Research}, vol. 12, no. Jul, pp.
  2415--2435, 2011.

\bibitem{fang2011sparse}
Z.~Fang,
\newblock ``Sparse group selection through co-adaptive penalties,''
\newblock {\em arXiv preprint arXiv:1111.4416}, 2011.

\bibitem{lounici2011oracle}
K.~Lounici, M.~Pontil, S.~Van~De Geer, and A.~B. Tsybakov,
\newblock ``Oracle inequalities and optimal inference under group sparsity,''
\newblock {\em The Annals of Statistics}, pp. 2164--2204, 2011.

\bibitem{vaiter2012degrees}
S.~Vaiter, C.~Deledalle, G.~Peyr{\'e}, J.~Fadili, and C.~Dossal,
\newblock ``The degrees of freedom of the group lasso for a general design,''
\newblock {\em arXiv preprint arXiv:1212.6478}, 2012.

\bibitem{bajwa2015conditioning}
W.~U. Bajwa, M.~F. Duarte, and R.~Calderbank,
\newblock ``Conditioning of random block subdictionaries with applications to
  block-sparse recovery and regression,''
\newblock {\em IEEE Transactions on Information Theory}, vol. 61, no. 7, pp.
  4060--4079, 2015.

\bibitem{ahsen2017}
M.~E. Ahsen and M.~Vidyasagar,
\newblock ``Error bounds for compressed sensing algorithms with group sparsity:
  {A} unified approach,''
\newblock {\em Applied and Computational Harmonic Analysis}, vol. 43, no. 2,
  2017.

\bibitem{cotter2005sparse}
S.~F. Cotter, B.~D. Rao, K.~Engan, and K.~Kreutz-Delgado,
\newblock ``Sparse solutions to linear inverse problems with multiple
  measurement vectors,''
\newblock {\em IEEE Transactions on Signal Processing}, vol. 53, no. 7, 2005.

\bibitem{tropp2006algorithms1}
J.~A. Tropp, A.~C. Gilbert, and M.~J. Strauss,
\newblock ``Algorithms for simultaneous sparse approximation. {P}art {I}:
  {G}reedy pursuit,''
\newblock {\em Signal Processing}, vol. 86, no. 3, pp. 572--588, 2006.

\bibitem{gribonval2008atoms}
R.~Gribonval, H.~Rauhut, K.~Schnass, and P.~Vandergheynst,
\newblock ``Atoms of all channels, unite! {A}verage case analysis of
  multi-channel sparse recovery using greedy algorithms,''
\newblock {\em Journal of Fourier analysis and Applications}, vol. 14, no. 5-6,
  pp. 655--687, 2008.

\bibitem{stojnic2009reconstruction}
M.~Stojnic, F.~Parvaresh, and B.~Hassibi,
\newblock ``On the reconstruction of block-sparse signals with an optimal
  number of measurements,''
\newblock {\em IEEE Transactions on Signal Processing}, vol. 57, no. 8, pp.
  3075--3085, 2009.

\bibitem{eldar2009robust}
Y.~C. Eldar and M.~Mishali,
\newblock ``Robust recovery of signals from a structured union of subspaces,''
\newblock {\em IEEE Transactions on Information Theory}, vol. 55, no. 11, pp.
  5302--5316, 2009.

\bibitem{eldar2010average}
Y.~C. Eldar and H.~Rauhut,
\newblock ``Average case analysis of multichannel sparse recovery using convex
  relaxation,''
\newblock {\em IEEE Transactions on Information Theory}, vol. 56, no. 1, pp.
  505--519, 2010.

\bibitem{eldar2010block}
Y.~C. Eldar, P.~Kuppinger, and H.~B{\"o}lcskei,
\newblock ``Block-sparse signals: Uncertainty relations and efficient
  recovery,''
\newblock {\em IEEE Transactions on Signal Processing}, vol. 58, no. 6, pp.
  3042--3054, 2010.

\bibitem{baraniuk2010model}
R.~G. Baraniuk, V.~Cevher, M.~F. Duarte, and C.~Hegde,
\newblock ``Model-based compressive sensing,''
\newblock {\em IEEE Transactions on Information Theory}, vol. 56, no. 4, pp.
  1982--2001, 2010.

\bibitem{stojnic2010ell}
M.~Stojnic,
\newblock ``$\ell_2/\ell_1$-optimization in block-sparse compressed sensing and
  its strong thresholds,''
\newblock {\em IEEE Journal of Selected Topics in Signal Processing}, vol. 4,
  no. 2, pp. 350--357, 2010.

\bibitem{boufounos2011sparse}
P.~Boufounos, G.~Kutyniok, and H.~Rauhut,
\newblock ``Sparse recovery from combined fusion frame measurements,''
\newblock {\em IEEE Transactions on Information Theory}, vol. 57, no. 6, pp.
  3864--3876, 2011.

\bibitem{fang2011recovery}
J.~Fang and H.~Li,
\newblock ``Recovery of block-sparse representations from noisy observations
  via orthogonal matching pursuit,''
\newblock {\em arXiv preprint arXiv:1109.5430}, 2011.

\bibitem{ben2011near}
Z.~Ben-Haim and Y.~C. Eldar,
\newblock ``Near-oracle performance of greedy block-sparse estimation
  techniques from noisy measurements,''
\newblock {\em IEEE Journal of Selected Topics in Signal Processing}, vol. 5,
  no. 5, 2011.

\bibitem{kim2012compressive}
J.~M. Kim, O.~K. Lee, and J.~C. Ye,
\newblock ``Compressive music: Revisiting the link between compressive sensing
  and array signal processing,''
\newblock {\em IEEE Transactions on Information Theory}, vol. 58, no. 1, pp.
  278--301, 2012.

\bibitem{davies2012rank}
M.~E. Davies and Y.~C. Eldar,
\newblock ``Rank awareness in joint sparse recovery,''
\newblock {\em IEEE Transactions on Information Theory}, vol. 58, no. 2, 2012.

\bibitem{lee2012subspace}
K.~Lee, Y.~Bresler, and M.~Junge,
\newblock ``Subspace methods for joint sparse recovery,''
\newblock {\em IEEE Transactions on Information Theory}, vol. 58, 2012.

\bibitem{elhamifar2012block}
E.~Elhamifar and R.~Vidal,
\newblock ``Block-sparse recovery via convex optimization,''
\newblock {\em IEEE Transactions on Signal Processing}, vol. 60, 2012.

\bibitem{rao2012universal}
N.~S. Rao, B.~Recht, and R.~D. Nowak,
\newblock ``Universal measurement bounds for structured sparse signal
  recovery,''
\newblock in {\em International Conference on Artificial Intelligence and
  Statistics}, 2012, pp. 942--950.

\bibitem{duarte2013measurement}
M.~F. Duarte, M.~B. Wakin, D.~Baron, S.~Sarvotham, and R.~G. Baraniuk,
\newblock ``Measurement bounds for sparse signal ensembles via graphical
  models,''
\newblock {\em IEEE Transactions on Information Theory}, vol. 59, 2013.

\bibitem{lv2011group}
X.~Lv, G.~Bi, and C.~Wan,
\newblock ``The group {L}asso for stable recovery of block-sparse signal
  representations,''
\newblock {\em IEEE Transactions on Signal Processing}, vol. 59, no. 4, pp.
  1371--1382, 2011.

\bibitem{druce2015anomaly}
J.~Druce, J.~D. Haupt, and S.~Gonella,
\newblock ``Anomaly-sensitive dictionary learning for structural diagnostics
  from ultrasonic wavefields,''
\newblock {\em IEEE Trans. Ultrasonics, Ferroelectrics, Freq. Control}, vol.
  62, no. 7, 2015.

\bibitem{druce2015structural}
J.~Druce, M.~Kadkhodaie, J.~D. Haupt, and S.~Gonella,
\newblock ``Structural diagnostics via anomaly-driven demixing of wavefield
  data,''
\newblock {\em International Workshop on Structural Health Monitoring (IWSHM)},
  2015.

\bibitem{druce2016defect}
J.~Druce, S.~Gonella, M.~Kadkhodaie, S.~Jain, and J.~D. Haupt,
\newblock ``Defect triangulation via demixing algorithms based on dictionaries
  with different morphological complexity,''
\newblock {\em Proceedings of 8th European Workshop on Structural Health
  Monitoring (IWSHM)}, 2016.

\bibitem{druce2017locating}
J.~Druce, S.~Gonella, M.~Kadkhodaie, S.~Jain, and J.~D. Haupt,
\newblock ``Locating material defects via wavefield demixing with
  morphologically germane dictionaries,''
\newblock {\em Structural Health Monitoring}, vol. 16, no. 1, 2017.

\bibitem{kadkhodaie2015locating}
M.~Kadkhodaie, S.~Jain, J.~D. Haupt, J.~Druce, and S.~Gonella,
\newblock ``Locating rare and weak material anomalies by convex demixing of
  propagating wavefields,''
\newblock in {\em 6th IEEE International Workshop on Computational Advances in
  Multi-Sensor Adaptive Processing (CAMSAP)}, 2015.

\bibitem{elyaderani2017group}
M.~K. Elyaderani, S.~Jain, J.~Druce, S.~Gonella, and J.~Haupt,
\newblock ``Group-level support recovery guarantees for group {L}asso
  estimator,''
\newblock in {\em IEEE International Conference on Acoustics, Speech and Signal
  Processing (ICASSP)}. IEEE, 2017, pp. 4366--4370.

\bibitem{bandeira2013certifying}
A.~S. Bandeira, E.~Dobriban, D.~G. Mixon, and W.~F. Sawin,
\newblock ``Certifying the restricted isometry property is hard,''
\newblock {\em IEEE Transactions on Information Theory}, vol. 59, no. 6, pp.
  3448--3450, 2013.

\bibitem{calderbank2015block}
R.~Calderbank, A.~Thompson, and Y.~Xie,
\newblock ``On block coherence of frames,''
\newblock {\em Applied and Computational Harmonic Analysis}, vol. 38, no. 1,
  pp. 50--71, 2015.

\bibitem{Sharma-et-al_Damage-Index_AIAA_2006}
V.~Sharma, S.~Hanagud, and M.~Ruzzene,
\newblock ``Damage index estimation in beams and plates using laser
  vibrometry,''
\newblock {\em AIAA Journal}, vol. 44, 2006.

\bibitem{Michaels_Ruzzene_Michaels_Ultrasonics_2010}
T.E. Michaels, J.E. Michaels, and M.~Ruzzene,
\newblock ``Frequency-wavenumber domain analysis of guided wavefields,''
\newblock {\em Ultrasonics}, vol. 51, no. 4, 2011.

\bibitem{Chandola09}
V.~Chandola, A.~Banerjee, and V.~Kumar,
\newblock ``Anomaly detection: {A} survey,''
\newblock {\em ACM Computing Surveys}, vol. 41, no. 3, 2009.

\bibitem{donoho2001uncertainty}
D.~L. Donoho and X.~Huo,
\newblock ``Uncertainty principles and ideal atomic decomposition,''
\newblock {\em IEEE Transactions on Information Theory}, vol. 47, no. 7, pp.
  2845--2862, 2001.

\bibitem{elad2005simultaneous}
M.~Elad, J.~L. Starck, P.~Querre, and D.~L. Donoho,
\newblock ``Simultaneous cartoon and texture image inpainting using
  morphological component analysis (mca),''
\newblock {\em Applied and Computational Harmonic Analysis}, vol. 19, no. 3,
  pp. 340--358, 2005.

\bibitem{amelunxen2014living}
D.~Amelunxen, M.~Lotz, M.~B. McCoy, and J.~A. Tropp,
\newblock ``Living on the edge: Phase transitions in convex programs with
  random data,''
\newblock {\em Information and Inference: A Journal of the IMA}, vol. 3, 2014.

\bibitem{foygel2014corrupted}
R.~Foygel and L.~Mackey,
\newblock ``Corrupted sensing: Novel guarantees for separating structured
  signals,''
\newblock {\em IEEE Transactions on Information Theory}, vol. 60, no. 2, pp.
  1223--1247, 2014.

\bibitem{levine2014block}
R.~Levine and J.~E. Michaels,
\newblock ``Block-sparse reconstruction and imaging for lamb wave structural
  health monitoring,''
\newblock {\em IEEE Transactions on Ultrasonics, Ferroelectrics, and Frequency
  Control}, vol. 61, 2014.

\bibitem{golato2016multimodal}
A.~Golato, S.~Santhanam, F.~Ahmad, and M.~G. Amin,
\newblock ``Multimodal sparse reconstruction in guided wave imaging of defects
  in plates,''
\newblock {\em Journal of Electronic Imaging}, vol. 25, no. 4, pp.
  043013--043013, 2016.

\bibitem{donoho1989uncertainty}
D.~L. Donoho and P.~B. Stark,
\newblock ``Uncertainty principles and signal recovery,''
\newblock {\em SIAM Journal on Applied Mathematics}, vol. 49, 1989.

\bibitem{kadkhodaie2014linear}
M.~Kadkhodaie, M.~Sanjabi, and Z-Q Luo,
\newblock ``On the linear convergence of the approximate proximal splitting
  method for non-smooth convex optimization,''
\newblock {\em Journal of the Operations Research Society of China}, vol. 2,
  no. 2, pp. 123--141, 2014.

\bibitem{wainwright2009sharp}
M.~J. Wainwright,
\newblock ``Sharp thresholds for high-dimensional and noisy sparsity recovery
  using $\ell_1$-constrained quadratic programming ({L}asso),''
\newblock {\em IEEE Transactions on Information Theory}, vol. 55, no. 5, 2009.

\bibitem{ravikumar2009sparse}
P.~Ravikumar, J.~Lafferty, H.~Liu, and L.~Wasserman,
\newblock ``Sparse additive models,''
\newblock {\em Journal of the Royal Statistical Society: Series B (Statistical
  Methodology)}, vol. 71, no. 5, pp. 1009--1030, 2009.

\bibitem{rudelson2013hanson}
M.~Rudelson and R.~Vershynin,
\newblock ``Hanson-{W}right inequality and sub-gaussian concentration,''
\newblock {\em Electron. Commun. Probab}, vol. 18, 2013.

\end{thebibliography}

\end{document}